\documentclass[aps,prd,twocolumn,superscriptaddress,nofootinbib,floatfix]{revtex4-2}
\usepackage{amsmath,amssymb,graphicx,bm}
\usepackage[colorlinks=true,linkcolor=blue,citecolor=blue,urlcolor=blue]{hyperref}
\begin{document}

\title{Impact of nonlinear electrodynamics on quasinormal modes of rotating black strings}

\author{F. S. Mirekhtiary}
\email{fatemeh.mirekhtiary@neu.edu.tr}
\affiliation{Faculty of Engineering, Near East University, Nicosia, North Cyprus via Mersin 10, Turkey}

\author{\.{I}. Sakall{\i}}
\email{izzet.sakalli@emu.edu.tr}
\affiliation{Physics Department, Eastern Mediterranean University, Famagusta 99628, North Cyprus via Mersin 10, Turkey}

\begin{abstract}
We study massless scalar perturbations of rotating anti-de Sitter black strings sourced by a power-Maxwell field with negative exponent $s$. The charge enters the metric function as $-Qr^{p}$ with $p=2/(1-2s)$, and the case $s=-1/2$ gives a linear term with an algebraic horizon. Since rotation is a local boost of the static string, the rotating spectrum follows exactly from the static one evaluated at a frequency-dependent momentum. We compute the spectrum with the Horowitz-Hubeny series, Chebyshev collocation and time-domain evolution, each implemented in two independent codes and checked against the planar Schwarzschild-AdS limit. The charge acts in two ways. At the reduced charge $\hat q_c=0.72626$ the least damped pair of modes meets on the imaginary axis, so more strongly charged strings ring down without oscillation, and at the endpoint of the branch the spectrum is exactly hypergeometric. At large transverse momentum the distance of the modes from the light cone grows as $k^{1/(1-4s)}$, replacing the $k^{-1/5}$ law of the neutral brane. Rotation continues the exceptional point into a line in the plane of charge and rotation, adds a purely damped axisymmetric tower, and never produces superradiant growth. A near-horizon estimate used earlier for this geometry misses all of these effects.
\end{abstract}

\maketitle

\section{Introduction}\label{isec1}

Quasinormal modes (QNMs) are the characteristic damped oscillations of a perturbed black hole. They control the ringdown of compact objects, they diagnose linear stability, and in asymptotically anti-de Sitter (AdS) spacetimes they are the poles of retarded correlators of the dual field theory at finite temperature \cite{Kokkotas:1999bd,Berti:2009kk,Konoplya:2011qq,Birmingham:2001pj,Son:2002sd,Kovtun:2005ev}. For AdS black holes with planar, toroidal or cylindrical horizons the scalar, electromagnetic and gravitational spectra are well charted \cite{Horowitz:1999jd,Cardoso:2001vs,Starinets:2002br,Morgan:2009vg,Morgan:2009pn}, and a recent line of work has turned to the fine structure of these spectra: overdamped and purely imaginary branches \cite{Jansen:2017oag,Grozdanov:2019uhi}, spectral bifurcations \cite{Hegde:2026}, pseudospectral instability \cite{Jaramillo:2020tuu,Arean:2023ejh,Cownden:2023dam,GarciaFarina:2026}, subexponential late-time decay \cite{Crump:2026}, second-order modes \cite{Pan:2025}, and signatures of thermodynamic phase transitions \cite{He:2010zb,Liu:2014gvf,Guo:2024}.

Black strings in AdS are the simplest rotating members of this family. Lemos found the cylindrical black hole of general relativity with a negative cosmological constant \cite{Lemos:1994xp}, and Lemos and Zanchin generated its rotating and charged version by a coordinate transformation that mixes time with the compact angle \cite{Lemos:1995cm}. The transformation is a boost, which is why the rotating string is locally static. The electromagnetic QNMs of rotating Lemos strings and their meaning for a moving plasma were obtained by Morgan, Miranda and Zanchin and by Mamani \textit{et al.} \cite{Morgan:2013dv,Mamani:2018qzl}. Superradiance of rotating black branes and strings was treated by Cardoso and Yoshida \cite{Cardoso:2005vk}, and Gregory-Laflamme-type questions for strings were followed through QNMs in Refs.~\cite{Gregory:1993vy,Konoplya:2008yy,Liu:2008ds}. Related analyses cover strings in massive gravity \cite{Ponglertsakul:2018smo}, strings dressed by thin shells \cite{Lin:2023}, three-dimensional strings and strong cosmic censorship \cite{Ho:2022}, and the large-$D$ limit \cite{Emparan:2014aba}.

Nonlinear electrodynamics (NLED) adds one more ingredient. Since the work of Born and Infeld \cite{Born:1934gh} and the conformally invariant power-Maxwell source of Hassa\"{\i}ne and Mart\'{\i}nez \cite{Hassaine:2007py}, NLED sources have been used to build static and rotating black strings \cite{Hendi:2010kv,Hendi:2013dwa,Bakhtiarizadeh:2024,Barbosa:2025,Alencar:2026reg} and slowly rotating toroidal holes \cite{Panotopoulos:2019}. Their scalar QNMs have been studied for Born-Infeld AdS holes \cite{Liu:2011bi}, for Einstein-power-Maxwell spacetimes \cite{Liu:2012epm}, for nonlinearly charged Lifshitz branes \cite{Gonzalez:2016}, and for Coulomb-like AdS holes in three dimensions \cite{Aragon:2021}. For the rotating power-Maxwell string the existing perturbative studies rest on tunneling arguments and on a near-horizon estimate of the spectrum \cite{Mirekhtiary:2024rbs,Mirekhtiary:2023qnm}. That estimate fixes the damping by the surface gravity alone. It therefore carries no information about the AdS boundary, where the NLED term of the metric function decays slowest, and it cannot see a change in the qualitative character of the spectrum.

Two features of the AdS problem make such a test necessary. The first is that the QNMs of planar AdS holes are boundary-value eigenmodes of the whole region outside the horizon: the Dirichlet wall at the boundary confines the field, and the frequencies are fixed by the interplay of the horizon with the near-boundary part of the potential \cite{Horowitz:1999jd,Kovtun:2005ev}. Large-momentum modes make this explicit, since they are trapped close to the boundary and their approach to the light cone is controlled by the leading correction to the AdS metric \cite{Festuccia:2008zx,Morgan:2009pn}. The second is that AdS spectra can change character under a smooth deformation, through collisions of modes on the imaginary axis and the exceptional points that accompany them \cite{Jansen:2017oag,Grozdanov:2019uhi,Hegde:2026}, and such collisions are where the spectrum is most sensitive to perturbations of the operator \cite{Jaramillo:2020tuu,Cownden:2023dam}. A power-Maxwell charge whose contribution to the metric function grows toward the boundary touches both features at once, and a criterion based on the surface gravity alone cannot detect either. Accurate spectra, cross-checked by independent methods, are therefore the minimum required to state what the charge does.

The present work replaces that estimate by the exact spectrum and asks a single question: what does the power-Maxwell charge change in the scalar ringdown of a rotating AdS black string, and which part of the answer is fixed by kinematics alone? Three statements answer it. First, rotation enters through an exact boost map, so the rotating spectrum is a known function of the static one, and the Hawking temperature of the rotating string sets its natural unit. Second, the charge drives an exceptional point of the static spectrum at $\hat q_c=0.72626$, beyond which the scalar ringdown is overdamped, and the endpoint of the branch is exactly solvable. Rotation continues this point into a line of exceptional points in the plane of charge and rotation. Third, the charge term dominates the approach to the AdS boundary and changes the large-momentum law of the modes from the $k^{-1/5}$ of the neutral brane \cite{Festuccia:2008zx,Morgan:2009pn} to $k^{1/(1-4s)}$. The gap we close is that neither the boost structure nor the near-boundary origin of these effects has been used for NLED strings, so the existing near-horizon estimate has not been tested against any benchmark.

The paper is organized as follows. Section~\ref{isec2} reviews the power-Maxwell black string, fixes the physical branch and the scaling variables, and records the boost that generates rotation. Section~\ref{isec3} derives the wave equation, the boost map and the boundary conditions, and describes the three numerical methods with their benchmarks. Section~\ref{isec4} presents the static spectrum, the exceptional point and the exact endpoint. Section~\ref{isec5} derives and tests the large-momentum law. Section~\ref{isec6} treats rotation. Section~\ref{isec7} collects the consistency checks and states the scope of the results, and Sec.~\ref{isec8} concludes. We use the signature $(-,+,+,+)$ and units $G=c=\hbar=1$.

\section{Power-Maxwell black strings}\label{isec2}

The background is a solution of Einstein gravity with a negative cosmological constant coupled to a power of the Maxwell invariant. We first recall the static solution and its branches, then introduce the variables that make the spectrum depend on a single charge parameter, and finally add rotation.

The action is
\begin{equation}
I=\frac{1}{16\pi}\int d^4x\sqrt{-g}\,\Big[R-2\Lambda-(-\mathcal F)^s\Big],\qquad \mathcal F=F_{\mu\nu}F^{\mu\nu},
\label{eq:action}
\end{equation}
with $\Lambda=-3/l^2$ and the sign convention of Ref.~\cite{Hendi:2010kv}, which keeps the Lagrangian real for an electric field. For the static string
\begin{equation}
ds^2=-f(r)\,dt^2+\frac{dr^2}{f(r)}+r^2d\phi^2+\frac{r^2}{l^2}\,dz^2,
\label{eq:static}
\end{equation}
the Maxwell equation $\partial_r(\sqrt{-g}\,\mathcal L_{\mathcal F}F^{rt})=0$ gives $F_{tr}=-q\lambda r^{\lambda-1}$ with $\lambda=(2s-3)/(2s-1)$, and the Einstein equations give
\begin{equation}
f(r)=\frac{r^2}{l^2}-\frac{M}{r}+C_s\,q^{2s}\,r^{\lambda-1},\qquad C_s=2^{s-1}(2s-1)\lambda^{2s-1},
\label{eq:fPMI}
\end{equation}
for $s<0$ and $0<s<1/2$, while for $s>1/2$ the sign of the last term is reversed \cite{Hendi:2010kv}. We have verified Eq.~\eqref{eq:fPMI} by inserting it into all components of the field equations, symbolically and at random rational points, for $s=-2,-1,-1/2,-1/4$ and $1/4$; the residuals vanish identically. The exponent of the charge term is
\begin{equation}
\lambda-1=-\frac{2}{2s-1},
\label{eq:exponent}
\end{equation}
so the term is linear in $r$ only for $s=-1/2$. A linear profile imposed for other values of $s$ does not solve the field equations; for $s=-1$, $-2$ and $1/4$ the residuals of that ansatz range from $0.1$ to $10^2$ at the same sample points.

The branch $0<s<1/2$ has $\lambda-1>2$, so the charge term overtakes the cosmological term and the geometry is not asymptotically AdS. We therefore work with $s<0$, where $C_s<0$ and
\begin{equation}
\begin{gathered}
f(r)=\frac{r^2}{l^2}-Q\,r^{p}-\frac{M}{r},\\ p=\frac{2}{1-2s}\in(0,2),\qquad Q=|C_s|q^{2s}>0 .
\end{gathered}
\label{eq:fp}
\end{equation}
The charge term grows more slowly than $r^2$ but faster than any term of the neutral string, and it is the leading correction to the AdS behavior at large $r$. The reference case is $s=-1/2$, $p=1$, with
\begin{equation}
f(r)=\frac{r^2}{l^2}-Q r-\frac{M}{r},\qquad Q=\frac{1}{4\sqrt2\,q}.
\label{eq:fRBS1}
\end{equation}
We call this the RBS1 profile. Its horizon is the unique positive root of $r^3-Ql^2r^2-Ml^2=0$; by Descartes' rule there is exactly one for $M>0$.

The spectrum depends on the parameters only through one ratio. Fixing the horizon radius $r_H$ and writing $x=r/r_H$, $u=r_H/r$, the horizon condition gives $M=r_H^3(1-\hat q)/l^2$ with
\begin{equation}
\begin{gathered}
\hat q\equiv\frac{Ql^2}{r_H^{2-p}},\qquad f=\frac{r_H^2}{l^2}\,\frac{h(u)}{u^2},\\ h(u)=1-\hat q\,u^{2-p}-(1-\hat q)\,u^3 .
\end{gathered}
\label{eq:qhat}
\end{equation}
Positivity of the mass requires $\hat q<1$, and the sign $C_s<0$ of Eq.~\eqref{eq:fPMI} with $q>0$ gives $\hat q>0$, so the physical range is $0\le\hat q<1$. The endpoint $\hat q=1$ is the string supported by the electric field alone, with $f=r(r-r_H)/l^2$ for $p=1$. The Hawking temperature of the static string is
\begin{equation}
T_s=\frac{f'(r_H)}{4\pi}=\frac{r_H}{4\pi l^2}\Big[3-(1+p)\hat q\Big],
\label{eq:Ts}
\end{equation}
which stays positive on the whole range because $(1+p)\hat q<3$. There is no extremal limit, so no near-extremal regime exists in this family, in agreement with the direct check that $f(r_e)=f'(r_e)=0$ requires $M<0$ when $Q>0$.

Rotation is introduced by the transformation of Ref.~\cite{Lemos:1995cm},
\begin{equation}
\mathcal T=\gamma t-a\phi,\qquad \Phi=\gamma\phi-\frac{a}{l^2}t,\qquad \gamma=\sqrt{1+\frac{a^2}{l^2}},
\label{eq:boost}
\end{equation}
which turns Eq.~\eqref{eq:static} into
\begin{equation}
ds^2=-f\,(\gamma dt-a\,d\phi)^2+r^2\Big(\frac{a}{l^2}dt-\gamma d\phi\Big)^2+\frac{dr^2}{f}+\frac{r^2}{l^2}dz^2 .
\label{eq:rot}
\end{equation}
Because $\phi$ is periodic while $\Phi$ in the static metric is not identified, Eq.~\eqref{eq:rot} is a globally distinct spacetime, but every local quantity is inherited from the static string. We find $\sqrt{-g}=r^2/l$, $g_{tt}=a^2r_H^2/l^4>0$ on the horizon, so an ergoregion surrounds the horizon, and the horizon angular velocity measured from the nonrotating boundary frame is
\begin{equation}
\Omega_H=\frac{a}{\gamma l^2},\qquad \Omega_H\,l=\frac{a/l}{\sqrt{1+a^2/l^2}}<1 .
\label{eq:OmegaH}
\end{equation}
The horizon generator is $\partial_t+\Omega_H\partial_\phi=\gamma^{-1}\partial_{\mathcal T}$, so the rotating string has temperature $T=T_s/\gamma$. The inequality in Eq.~\eqref{eq:OmegaH} is the Hawking-Reall condition \cite{Hawking:1999dp}, and it holds for every value of $a$.

\section{Scalar perturbations and methods}\label{isec3}

This section derives the radial equation on the rotating background, shows that it is the static equation in disguise, fixes the boundary conditions, and describes the three numerical routes together with the benchmarks they reproduce.

\subsection{Wave equation and boost map}

A test scalar obeys $\Box\Psi=\mu^2\Psi$. With $\Psi=e^{-i\omega t+im\phi+ikz}R(r)$ on Eq.~\eqref{eq:rot}, the inverse metric gives the radial equation
\begin{equation}
\frac{1}{r^2}\frac{d}{dr}\Big(r^2f\frac{dR}{dr}\Big)+\Big[\frac{\Omega^2}{f}-\frac{P^2}{r^2}-\mu^2\Big]R=0,
\label{eq:radial}
\end{equation}
with
\begin{equation}
\Omega=\gamma\omega-\frac{a m}{l^2},\qquad P^2=\mathcal M^2+l^2k^2,\qquad \mathcal M=\gamma m-a\omega .
\label{eq:map}
\end{equation}
We checked Eq.~\eqref{eq:radial} symbolically. The pair $(\Omega,\mathcal M)$ is the frequency and angular wave number in the static frame of Eq.~\eqref{eq:boost}, since $\omega t-m\phi=\Omega\mathcal T-\mathcal M\Phi$. Equation~\eqref{eq:radial} is thus the static radial equation with frequency $\Omega$ and transverse momentum $P$. If $\Omega_n(P^2)$ denotes the static spectrum, analytically continued to complex $P^2$, the rotating spectrum solves
\begin{equation}
\Omega=\Omega_n\big(P^2\big),\qquad \mathcal M=\frac{m-a\Omega}{\gamma},\qquad \omega=\frac{1}{\gamma}\Big(\Omega+\frac{a m}{l^2}\Big).
\label{eq:boostmap}
\end{equation}
The map is exact. It shows that rotation acts on the spectrum only through the factor $\gamma$ and through the frequency-dependent momentum $\mathcal M$, which is complex for a damped mode even when $m=0$. In units of the temperature of the rotating string the imaginary parts obey $\mathrm{Im}\,\omega/(2\pi T)=\mathrm{Im}\,\Omega/(2\pi T_s)$, so the factor $\gamma$ drops out and only the shift of $P^2$ remains. The superradiant condition $\mathrm{Re}\,\omega<m\Omega_H$ becomes $\mathrm{Re}\,\Omega<0$, which is the mirror branch of a static spectrum that is symmetric under $\Omega\to-\bar\Omega$ for real $P^2$.

\begin{figure*}[t]
\centering
\includegraphics[width=0.86\textwidth]{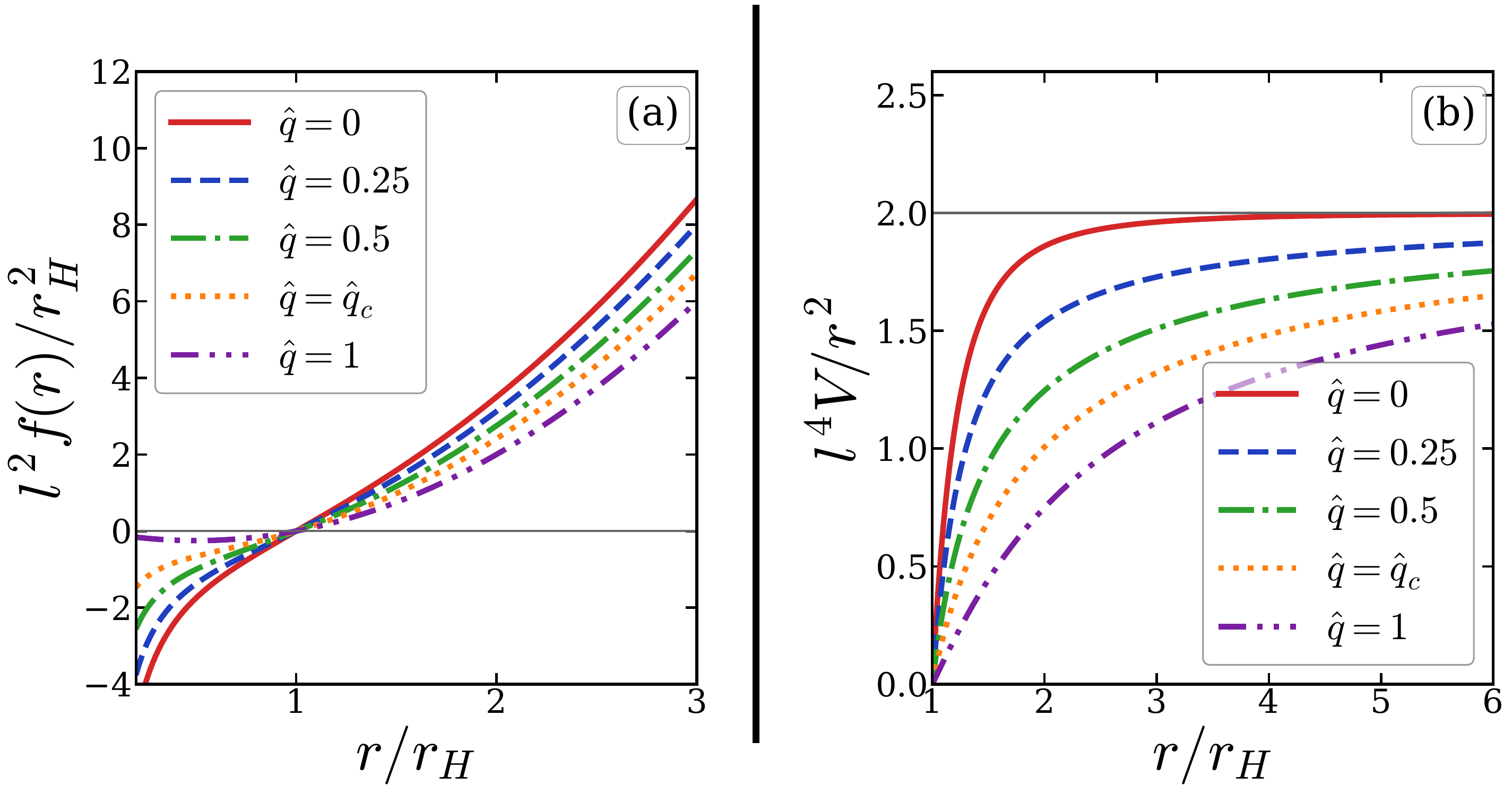}
\caption{(a) Metric function $l^2f/r_H^2$ of the RBS1 profile, Eq.~\eqref{eq:fRBS1}, for reduced charges $\hat q=0$, $0.25$, $0.5$, $\hat q_c$ and $1$. All curves share the horizon at $r=r_H$, and the mass term disappears at $\hat q=1$. (b) Reduced effective potential $l^4V/r^2$ of a massless scalar with $P=0$. The neutral string reaches the AdS value $2$ within a few horizon radii, while the charge term delays the approach, leaving a deficit that decays only as $\hat q\,r_H/r$.}
\label{fig:metric}
\end{figure*}

Introducing $\psi=rR$ and the tortoise coordinate $dr_*=dr/f$, normalized so that $r_*\to0$ at the boundary, Eq.~\eqref{eq:radial} becomes $\psi''+(\Omega^2-V)\psi=0$ with
\begin{equation}
V=f\Big(\mu^2+\frac{f'}{r}+\frac{P^2}{r^2}\Big).
\label{eq:V}
\end{equation}
Near the horizon $V\to0$ and the QNM condition is $\psi\sim e^{-i\Omega r_*}$. Near the boundary $V\simeq(2+\mu^2l^2)/r_*^2$ and the solutions behave as $R\sim r^{-\Delta_\pm}$ with $\Delta_\pm=\tfrac32\pm\sqrt{\tfrac94+\mu^2l^2}$; we impose the normalizable falloff $r^{-\Delta_+}$, which requires the Breitenlohner-Freedman bound $\mu^2l^2\ge-9/4$ \cite{Breitenlohner:1982jf}. From here on we take $\mu=0$, so $\Delta_+=3$; the scalar is then dual to a marginal operator of dimension three. An earlier treatment of this geometry identified $\mu^2$ with $L(L+1)$; the planar horizon carries no multipole index, and the angular and axial labels enter only through $P^2$. Figure~\ref{fig:metric} shows $f$ and the reduced potential $l^4V/r^2$, which interpolates between zero at the horizon and its AdS value two at the boundary. The charge slows this approach to $1-\hat q\,u+\dots$ instead of $1-u^3$; this slow approach is behind the large-momentum law of Sec.~\ref{isec5}.

\subsection{Three independent routes}

The radial equation is regular in ingoing Eddington-Finkelstein form. Setting $l=r_H=1$, $R=e^{-i\Omega r_*}\varphi(u)$ and $\mu=0$, Eq.~\eqref{eq:radial} becomes
\begin{equation}
u\,h\,\varphi''+\big(uh'-2h+2i\Omega u\big)\varphi'-\big(2i\Omega+P^2u\big)\varphi=0 .
\label{eq:EF}
\end{equation}
The first route is the Horowitz-Hubeny (HH) series \cite{Horowitz:1999jd}, $\varphi=\sum_j a_j(u-1)^j$ with $a_0=1$. For $p=1$ the coefficients of Eq.~\eqref{eq:EF} are polynomials, the recurrence closes with five terms, and the QNM condition is $\varphi(0)=\sum_j a_j(-1)^j=0$. The radius of convergence equals one for every $\hat q<1$, because the remaining zeros of $h$, the roots of $(1-\hat q)u^2+(1-\hat q)u+1=0$, lie at distance larger than one from $u=1$. We evaluate the series with 300 to 600 terms in 30 to 60 digit arithmetic and accept a root only if it is stable under a 50\% increase of the truncation order. In the rotating case $\Omega$ and $P^2$ are polynomials in $\omega$ and the same recurrence applies \cite{Zhidenko:2005mv,Fernandes:2021,Lin:2023}.

The second route is Chebyshev collocation of Eq.~\eqref{eq:EF} for $\chi=u^{-3}\varphi$ on Gauss-Lobatto points in $u\in[0,1]$. The static problem is a generalized linear eigenvalue problem in $\Omega$; the rotating problem is quadratic in $\omega$ and we linearize it with a companion matrix. Spurious eigenvalues are removed by comparing two resolutions. For $p\neq1$ the term $u^{2-p}$ is not analytic at the boundary, and we restore spectral accuracy with the map $u=w^3$, which turns $u^{2-p}$ into a polynomial in $w$ for $s=-1$ and $s=-1/4$.

The third route evolves $\psi(t,r_*)$ in the time domain. For the static string and for the axisymmetric sector of the rotating string the evolution equation is
\begin{equation}
\Big(\gamma^2-\frac{a^2}{l^2}h\Big)\partial_t^2\psi=\partial_{r_*}^2\psi-V_0\,\psi,\qquad V_0=f\Big(\frac{f'}{r}+\frac{l^2k^2}{r^2}\Big),
\label{eq:TD}
\end{equation}
obtained from Eq.~\eqref{eq:radial} with $\omega\to i\partial_t$ and $m=0$. The prefactor is positive everywhere outside the horizon, so the problem is hyperbolic with wave speed one at the boundary and $\gamma^{-1}$ at the horizon. We use second-order finite differences on a uniform $r_*$ grid with $\Delta r_*=0.005$, a Dirichlet condition at the boundary and an outgoing condition at $r_*=-40$, in the spirit of Refs.~\cite{Gundlach:1993tp,Wang:2000gsa}, and extract frequencies with the matrix pencil method \cite{Hua:1990}.

Each route was implemented twice in independent codes, one of them written in Mathematica. Table~\ref{tab:bench} lists the benchmarks. The neutral planar value $\omega l^2/r_H=1.849421-2.663853i$ is the large-black-hole limit of the spherical values of Ref.~\cite{Horowitz:1999jd}, where $r_+=100$ gives $184.95-266.38i$; the endpoint values follow from the exact solution of Sec.~\ref{isec4}. The HH and collocation results agree to all digits shown, the two HH implementations agree to twelve digits, a 30-digit collocation reproduces the HH overtones to ten digits, and the time-domain values agree to three digits, which is the accuracy expected from a finite ringdown window.

\begin{table*}[t]
\caption{Benchmarks and cross-checks for $s=-1/2$ in units $l=r_H=1$, massless scalar, $k=0$. HH: Horowitz-Hubeny series; Cheb.: Chebyshev collocation; TD: time domain with matrix pencil; HH$'$: independent second HH implementation.}
\label{tab:bench}
\renewcommand{\arraystretch}{1.35}
\begin{ruledtabular}
\begin{tabular}{lll}
\textbf{Case} & \textbf{Route} & \textbf{$\omega\,l^2/r_H$}\\
\hline
$\hat q=0$, $n=0$ & HH, Cheb. & $1.849421-2.663853i$\\
 & TD & $1.8500-2.6674i$\\
 & HH$'$ & $1.849421313585-2.663852661153i$\\
$\hat q=0$, $n=1$ & HH & $3.161257-4.916418i$\\
$\hat q=0.5$, $n=0$ & HH, Cheb. & $0.775525-2.114101i$\\
 & TD & $0.7785-2.1138i$\\
 & HH$'$ & $0.775525048095-2.114101462034i$\\
$\hat q=0.9$, $n=0$ & HH, Cheb. & $-0.982037i$\\
 & TD & $-0.9828i$\\
$\hat q=1$, $n=0,1,2$ & exact & $-\tfrac34 i,\ -\tfrac43 i,\ -\tfrac{15}{8}i$\\
 & HH & $-0.750000i,\ -1.333333i,\ -1.875000i$\\
$\hat q=0.3$, $a=0.5$, $m=0$ & HH, HH$'$ & $1.168764-3.151509i$\\
 & HH & $-2.100235i$\\
$\hat q=0.3$, $a=2$, $m=0$ & HH, HH$'$ & $-0.675681i$\\
 & TD & $-0.6757i$\\
$s=-1$, $\hat q=0.5$ & Cheb. & $0.849942-2.323596i$\\
 & TD & $0.8520-2.3229i$\\
\end{tabular}
\end{ruledtabular}
\end{table*}

\section{Static spectrum, overdamping and the exact endpoint}\label{isec4}

We now follow the static RBS1 spectrum across the physical range of the charge. The neutral end is the planar SAdS$_4$ spectrum; the charged end turns out to be exactly solvable; and between the two the spectrum changes character at an exceptional point.

\begin{figure*}[t]
\centering
\includegraphics[width=\textwidth]{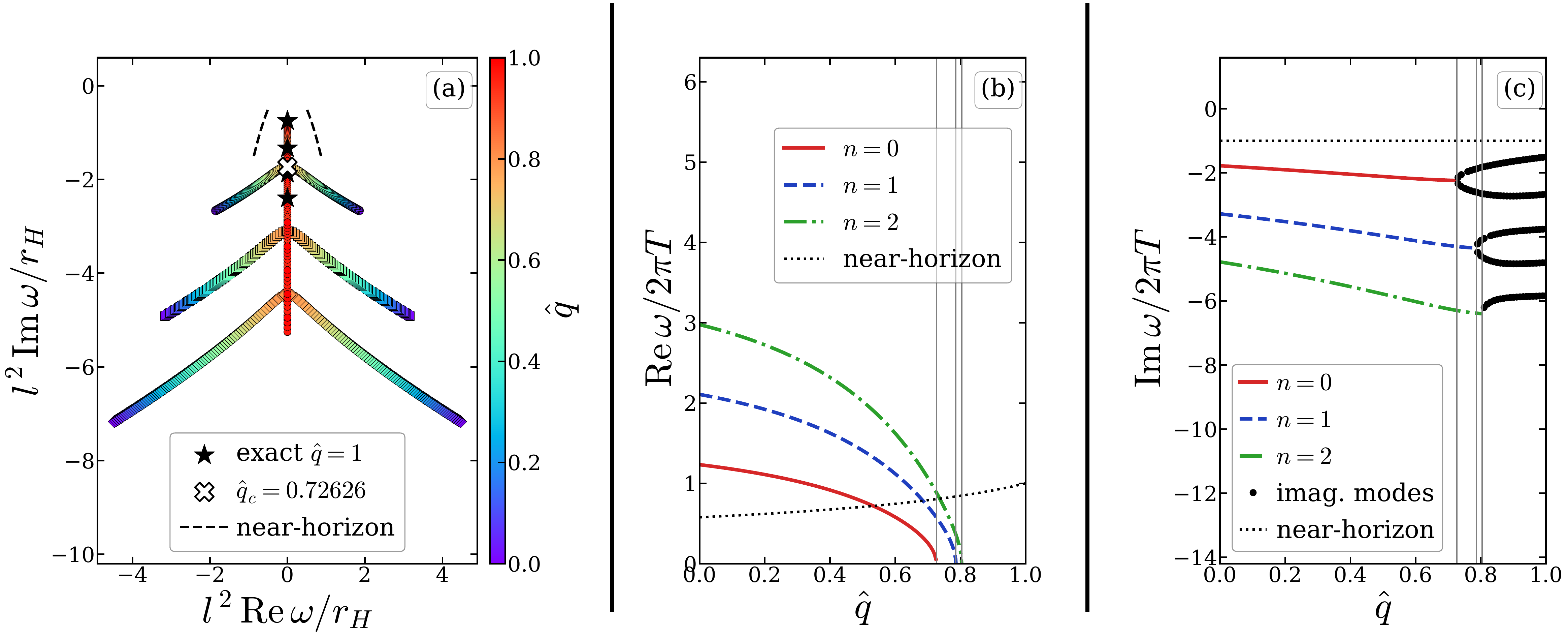}
\caption{Static scalar spectrum of the RBS1 string ($s=-1/2$, $k=0$). (a) Fundamental mode and first two overtones in the complex plane, colored by $\hat q$. The pairs move toward the imaginary axis and collide there, the fundamental at the exceptional point $\hat q_c$ (cross), after which purely imaginary modes run toward the exact values of the $\hat q=1$ string (stars). The near-horizon estimate (dashed) is far from every branch. (b) Real part and (c) imaginary part in units of $2\pi T_s$. Vertical lines mark the collisions of the three branches.}
\label{fig:spectrum}
\end{figure*}

The least damped modes and their dependence on $\hat q$ are listed in Table~\ref{tab:static} and drawn in Fig.~\ref{fig:spectrum}. Increasing the charge lowers both the oscillation frequency and the damping rate in units of $r_H/l^2$. Part of this trend is kinematic, because the temperature of Eq.~\eqref{eq:Ts} falls from $3r_H/(4\pi l^2)$ to $r_H/(4\pi l^2)$ across the range. Measured in units of $2\pi T_s$, the oscillation frequency still falls, whereas the damping rate first rises and reaches $|\mathrm{Im}\,\omega|/(2\pi T_s)\simeq2.24$ at $\hat q_c$, as panels (b) and (c) of Fig.~\ref{fig:spectrum} show. The charge therefore does not act through the temperature alone. Tracking the fundamental pair $\pm\mathrm{Re}\,\omega+i\,\mathrm{Im}\,\omega$ with the HH series, we find that the real part vanishes at
\begin{equation}
\begin{gathered}
\hat q_c=0.72625663,\\ \omega_c\,\frac{l^2}{r_H}=-1.7296186\,i,\qquad \frac{\omega_c}{2\pi T_s}=-2.2354\,i,
\end{gathered}
\label{eq:qc}
\end{equation}
obtained by solving the double-root conditions $\mathcal F(\omega)=\partial_\omega\mathcal F(\omega)=0$ for the HH characteristic function $\mathcal F$ on the imaginary axis; the two independent codes agree to eight digits. Close to this point $(\mathrm{Re}\,\omega)^2$ is linear in $\hat q_c-\hat q$, with $\mathrm{Re}\,\omega\,l^2/r_H\simeq1.22\,(\hat q_c-\hat q)^{1/2}$, which is the square-root signature of an exceptional point where two QNMs and their eigenfunctions coalesce. Beyond $\hat q_c$ the pair splits into two purely imaginary modes; one moves quickly toward the real axis, while the other first moves away from it and then turns back. The first and second overtones follow the same pattern at $\hat q\simeq0.786$ and $\hat q\simeq0.804$, so overdamping propagates up the tower as the mass term is switched off. Purely imaginary branches and bifurcations of this kind are known for Schwarzschild-de Sitter holes \cite{Jansen:2017oag}, for black branes \cite{Grozdanov:2019uhi} and for regular BTZ holes \cite{Hegde:2026}; here the parameter that drives them is the NLED charge.

\begin{table*}[t]
\caption{Static RBS1 spectrum ($s=-1/2$, $k=0$, $l=r_H=1$): the two least damped modes from the HH series, the fundamental in units of $2\pi T_s=(3-2\hat q)/2$, and the near-horizon estimate $\omega_{\rm NH}=\sqrt{\kappa/2}-i\kappa$ with $\kappa=2\pi T_s$.}
\label{tab:static}
\renewcommand{\arraystretch}{1.35}
\begin{ruledtabular}
\begin{tabular}{lllll}
\textbf{$\hat q$} & \textbf{$\omega_0\,l^2/r_H$} & \textbf{$\omega_1\,l^2/r_H$} & \textbf{$\omega_0/2\pi T_s$} & \textbf{$\omega_{\rm NH}$}\\
\hline
0    & $1.849421-2.663853i$ & $3.161257-4.916418i$ & $1.23295-1.77590i$ & $0.866-1.50i$\\
0.25 & $1.337210-2.419393i$ & $2.325860-4.483866i$ & $1.06977-1.93551i$ & $0.791-1.25i$\\
0.5  & $0.775525-2.114101i$ & $1.409569-3.959630i$ & $0.77553-2.11410i$ & $0.707-1.00i$\\
0.7  & $0.206577-1.782896i$ & $0.571096-3.412017i$ & $0.25822-2.22862i$ & $0.632-0.80i$\\
0.8  & $-1.282354i$ & $-1.845868i$ & $-1.83193i$ & $0.592-0.70i$\\
0.9  & $-0.982037i$ & $-1.633496i$ & $-1.63673i$ & $0.548-0.60i$\\
1    & $-0.750000i$ & $-1.333333i$ & $-1.50000i$ & $0.500-0.50i$\\
\end{tabular}
\end{ruledtabular}
\end{table*}

The endpoint $\hat q=1$ is solvable in closed form. With $f=r(r-r_H)/l^2$, $P=0$ and $z=r_H/r$, Eq.~\eqref{eq:radial} has regular singular points only at $z=0$, $1$ and $\infty$, with exponents $\{0,3\}$, $\{\pm i\hat\omega\}$ and $\{-1\pm\sqrt{1-\hat\omega^2}\}$, where $\hat\omega=\omega l^2/r_H$. The solution that is normalizable at the boundary and ingoing at the horizon is
\begin{equation}
\begin{gathered}
R=z^3(1-z)^{-i\hat\omega}\,{}_2F_1\big(\alpha,\beta;4;z\big),\\ \alpha,\beta=2\pm\sqrt{1-\hat\omega^2}-i\hat\omega ,
\end{gathered}
\label{eq:exactR}
\end{equation}
and the outgoing component at $z=1$ is proportional to $1/[\Gamma(\alpha)\Gamma(\beta)]$. It vanishes when $\beta=-n$, which gives the purely imaginary spectrum
\begin{equation}
\begin{gathered}
\omega_n=-i\,\frac{r_H}{l^2}\,\frac{(n+1)(n+3)}{2(n+2)},\\ \frac{\omega_n}{2\pi T_s}=-i\Big(n+2-\frac{1}{n+2}\Big),
\end{gathered}
\label{eq:exact}
\end{equation}
with $n=0,1,2,\dots$, the hypergeometric function then being a polynomial of degree $n$. The values $-3i/4$, $-4i/3$ and $-15i/8$ are reproduced to forty digits by the HH series and are the endpoints of the imaginary branches in Fig.~\ref{fig:spectrum}. In units of $2\pi T_s$ the spacing of Eq.~\eqref{eq:exact} tends to one, the Matsubara spacing, while the low modes are pushed below the Matsubara values by $1/(n+2)$.

The near-horizon estimate $\omega_{\rm NH}=\sqrt{\kappa/2}-i(2n+1)\kappa$, with $\kappa=f'(r_H)/2$, which follows from the approximation used earlier for this geometry at $P=0$, is listed in the last column of Table~\ref{tab:static}. Its damping rate is too small by a factor close to two at $\hat q=0$ and by a factor of $1.5$ at $\hat q=1$, in units of $2\pi T_s$ it is pinned at $-i$ for every charge, whereas the exact value moves from $-1.78i$ to $-2.24i$ and back, and it has no exceptional point. The reason is structural: the estimate keeps only the exponential tail of the potential near the horizon, whereas in AdS the modes are set by the whole region between the horizon and the boundary, and the charge modifies the potential most strongly near the boundary.

\section{Large-momentum law}\label{isec5}

The slow approach of the potential to its AdS form, visible in Fig.~\ref{fig:metric}(b), has a sharp consequence at large transverse momentum. We first derive the asymptotic law, then test it numerically for three values of $s$, and finally note which part of the result is universal.

At large $\hat k=l^2k/r_H$ the least damped modes approach the light cone and their wave functions are confined near the boundary. Let the metric function there be $l^2f/r^2=1-c\,u^{\nu}+\dots$, with $\nu=2-p$ and $c=\hat q$ for the charged string, and $\nu=3$ and $c=1$ for the neutral one. With $\zeta=-r_*r_H/l^2\simeq u$ the wave equation becomes, to leading order,
\begin{equation}
\psi''+\Big(\hat\omega^2-\hat k^2+c\,\hat k^2\zeta^{\nu}-\frac{2}{\zeta^2}\Big)\psi=0 .
\label{eq:nearbdy}
\end{equation}
The rescaling $\zeta=(c\hat k^2)^{-1/(\nu+2)}y$ and $\hat\omega^2-\hat k^2=-\varepsilon\,(c\hat k^2)^{2/(\nu+2)}$ removes all parameters,
\begin{equation}
\frac{d^2\psi}{dy^2}+\Big(y^{\nu}-\varepsilon-\frac{2}{y^2}\Big)\psi=0,
\label{eq:scaled}
\end{equation}
with $\psi\sim y^2$ at the boundary and an ingoing wave $\psi\sim\exp(\tfrac{2i}{\nu+2}y^{(\nu+2)/2})$ toward the horizon. On the ray $y=e^{i\pi/(\nu+2)}\rho$ the ingoing solution decays and Eq.~\eqref{eq:scaled} becomes the real eigenvalue problem
\begin{equation}
-\frac{d^2\psi}{d\rho^2}+\Big(\rho^{\nu}+\frac{2}{\rho^2}\Big)\psi=\lambda_j\psi,\qquad \varepsilon_j=\lambda_j\,e^{i\pi\nu/(\nu+2)} .
\label{eq:lambda}
\end{equation}
Expanding $\hat\omega$ in $\hat k$ gives the asymptotic law
\begin{equation}
\hat\omega_j\simeq\hat k-\frac{\lambda_j}{2}\,e^{i\pi\nu/(\nu+2)}\,c^{2/(\nu+2)}\,\hat k^{(2-\nu)/(\nu+2)} .
\label{eq:largek}
\end{equation}
For the neutral string, $\nu=3$ gives $\hat k^{-1/5}$ and the phase $e^{-2\pi i/5}$ of the deviation, in agreement with the Bohr-Sommerfeld analysis of Festuccia and Liu \cite{Festuccia:2008zx} and the numerical eikonal study of Morgan \textit{et al.} \cite{Morgan:2009pn}. For the power-Maxwell string $\nu=2-p$ is smaller than three for every $s<0$, so the charge term always dominates, and Eq.~\eqref{eq:largek} becomes
\begin{equation}
\hat\omega_j-\hat k\simeq-\frac{\lambda_j}{2}\,e^{-2\pi is/(1-4s)}\,\hat q^{(1-2s)/(1-4s)}\,\hat k^{1/(1-4s)} .
\label{eq:largekPMI}
\end{equation}
The exponent $1/(1-4s)$ is positive. The ratio $\mathrm{Re}\,\omega/k$ still tends to one, but $\omega-k$ and the damping grow without bound, so the long-lived high-momentum modes of the neutral brane disappear as soon as $\hat q>0$. For RBS1 the law reads $\hat k^{1/3}$ with phase $e^{i\pi/3}$, so that $\mathrm{Im}\,\omega/\mathrm{Re}(\omega-k)\to\sqrt3$.

\begin{figure*}[t]
\centering
\includegraphics[width=\textwidth]{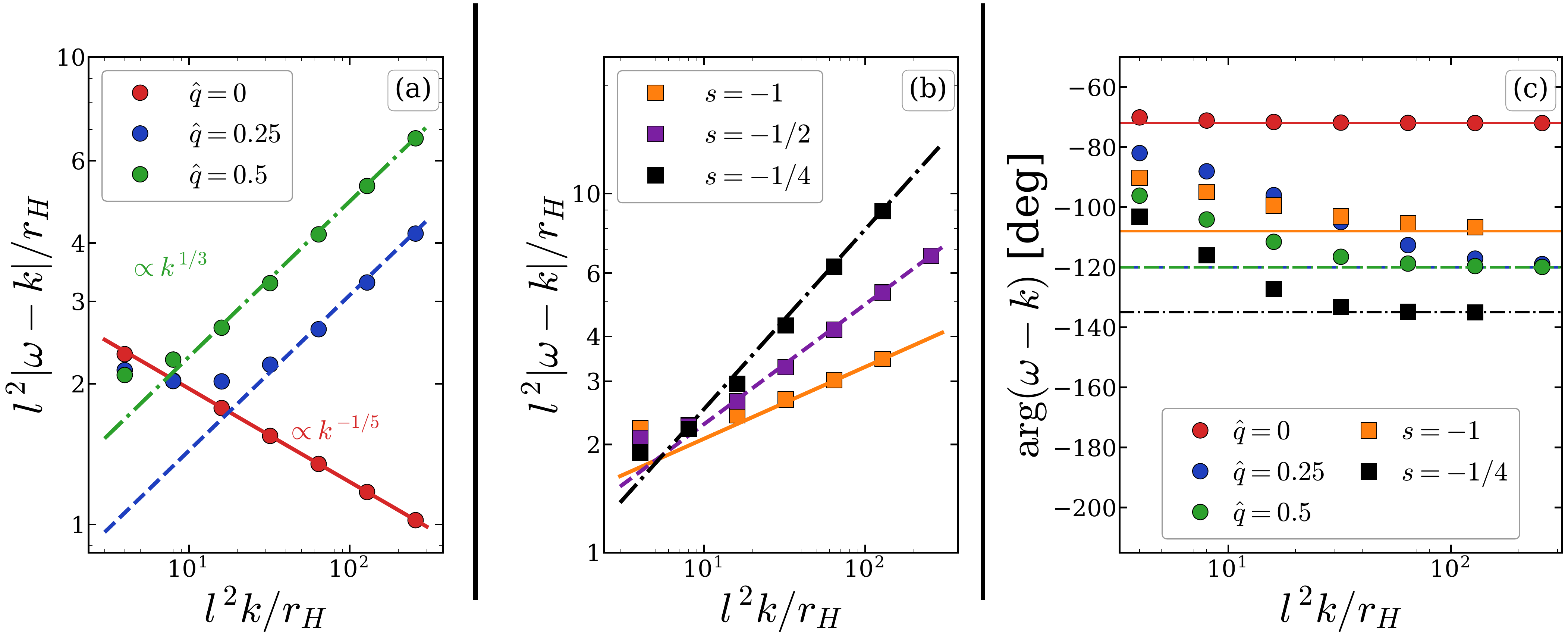}
\caption{Large-momentum behavior of the least damped mode. Symbols are collocation results, lines are the asymptotic law, Eq.~\eqref{eq:largek}, with no fitted parameter. (a) RBS1 ($s=-1/2$): the deviation from the light cone decreases as $\hat k^{-1/5}$ for the neutral string and grows as $\hat k^{1/3}$ once $\hat q>0$. (b) Charged strings with $\hat q=0.5$ for $s=-1$, $-1/2$ and $-1/4$, with slopes $1/5$, $1/3$ and $1/2$. (c) Phase of $\omega-k$ approaching $-72^\circ$, $-120^\circ$, $-108^\circ$ and $-135^\circ$.}
\label{fig:largek}
\end{figure*}

The eigenvalues of Eq.~\eqref{eq:lambda} were computed by finite differences and, independently, with a finite-element eigensolver. For $\nu=3$, $1$, $4/3$ and $2/3$ the lowest values are $\lambda_1=6.1918$, $3.3612$, $3.9715$ and $2.6746$, the two codes agreeing to four digits. Figure~\ref{fig:largek} compares Eq.~\eqref{eq:largek} with collocation spectra up to $\hat k=256$. For the neutral string at $\hat k=128$ the prediction is $0.3624-1.1157i$ against the computed $0.3631-1.1153i$. For RBS1 with $\hat q=0.5$ at $\hat k=256$ it is $-3.361-5.822i$ against $-3.339-5.809i$, and the exponents $1/5$ and $1/2$ for $s=-1$ and $s=-1/4$ are reproduced with the predicted phases of $-108^\circ$ and $-135^\circ$. At the largest momenta the differences are below $3\%$ for every case, and for RBS1 they fall from $1.0\%$ at $\hat k=128$ to $0.4\%$ at $\hat k=256$.

Equation~\eqref{eq:largek} separates two kinds of input. The numbers $\lambda_j$ and the phase depend only on the leading power $\nu$ of the near-boundary deviation, and are therefore universal for every planar AdS$_4$ background with a given $\nu$. The model enters through $\nu=2-p$ and through the coefficient $c=\hat q$. In holographic language the charge term changes the ultraviolet structure of the thermal correlator of the dual operator, because the electric field grows toward the boundary and the deformation it induces is felt at all momenta; the modification is not a property of the horizon.

\section{Rotation}\label{isec6}

We turn to the rotating string, where Eq.~\eqref{eq:boostmap} reduces every question to the static spectrum at complex momentum. We first discuss the axisymmetric sector, which carries a new branch, then the co-rotating and counter-rotating modes, and finally superradiance, including the line of exceptional points that connects rotation with the static result of Sec.~\ref{isec4}.

\begin{figure*}[t]
\centering
\includegraphics[width=\textwidth]{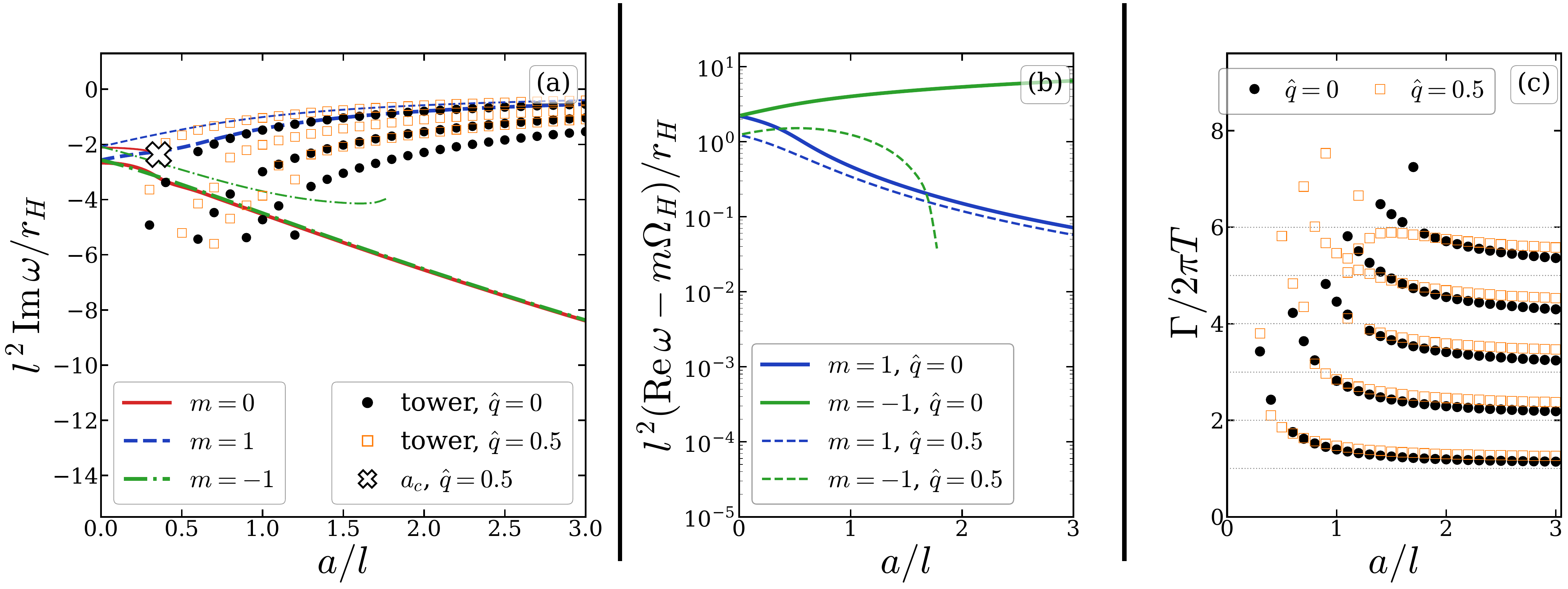}
\caption{Rotating RBS1 string, $k=0$; thick lines and filled symbols refer to $\hat q=0$, thin lines and open symbols to $\hat q=0.5$. (a) Imaginary parts of the complex fundamental modes with $m=0$ and $m=\pm1$ and of the lowest purely damped axisymmetric modes, for $\hat q=0$ and $\hat q=0.5$. The damped tower enters from $-i\infty$ and overtakes the complex $m=0$ mode at $a_\times$. (b) Distance $\mathrm{Re}\,\omega-m\Omega_H=\mathrm{Re}\,\Omega/\gamma$ of the $m=\pm1$ modes from the superradiant threshold. It is positive on every tracked branch; the co-rotating modes approach the threshold as $\Omega_H l\to1$, and the strongly damped counter-rotating mode of the charged string approaches it near $a/l\simeq1.78$. (c) The damped tower in units of $2\pi T=2\pi T_s/\gamma$, approaching a ladder with nearly unit spacing.}
\label{fig:rot}
\end{figure*}

\begin{figure*}[t]
\centering
\includegraphics[width=0.86\textwidth]{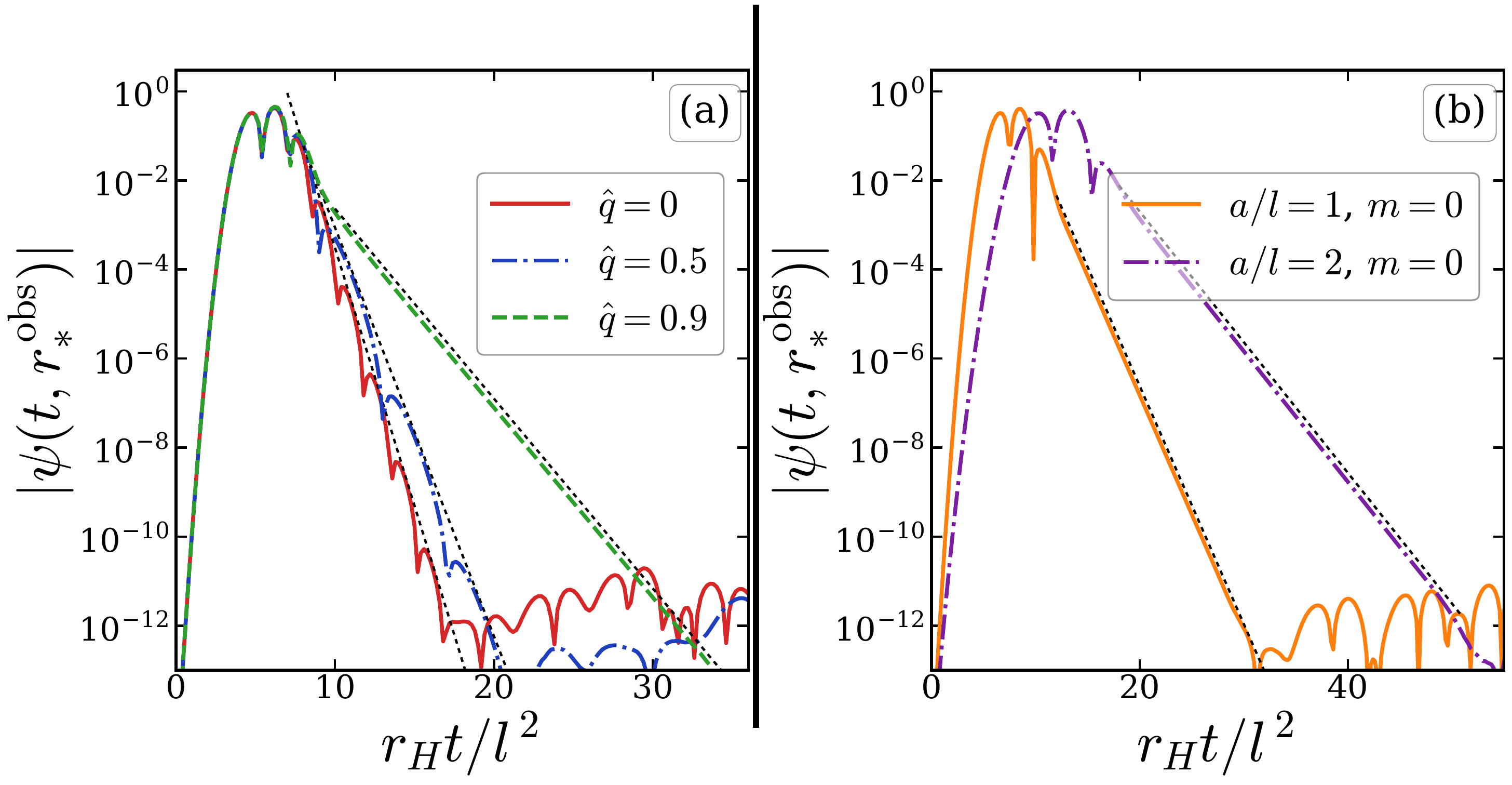}
\caption{Time-domain profiles at $r_*^{\rm obs}=-l^2/r_H$ after a Gaussian pulse. Dotted lines are the exponential envelopes of the leading frequency obtained with the matrix pencil method. (a) Static RBS1 string: oscillatory ringdown at $\hat q=0$ and $0.5$, and a non-oscillatory tail at $\hat q=0.9$, beyond the exceptional point, decaying at the rate $0.9828\,r_H/l^2$. (b) Axisymmetric perturbations of the rotating string with $\hat q=0.3$, where the purely damped tower dominates the late signal. The noise floor near $10^{-12}$ is set by the discretization.}
\label{fig:td}
\end{figure*}

For $m=0$ the static-frame data are $\Omega=\gamma\omega$ and $\mathcal M=-a\omega$, so $P^2=a^2\omega^2+l^2k^2$. A damped mode therefore probes the static spectrum at complex $P^2$, and a purely damped mode $\omega=-i\Gamma$ probes it at negative $P^2=-a^2\Gamma^2$, where the transverse term lowers the potential of Eq.~\eqref{eq:V}. This is the origin of a tower of purely imaginary axisymmetric modes that exists for every $a>0$. As $a\to0$ the tower recedes to $-i\infty$, consistent with its absence in the static string; as $a$ grows its members move up the imaginary axis, with the lowest one at $\Gamma l^2/r_H=3.38$ for $a/l=0.4$ and $0.58$ for $a/l=2.8$ in the neutral case. Figure~\ref{fig:rot}(a) shows that for $\hat q=0$ the lowest member overtakes the complex fundamental at $a_\times/l\simeq0.41$. For $\hat q=0.5$ something stronger happens: the complex $m=0$ pair itself reaches the imaginary axis, at $a_c/l=0.35673$ and $\Gamma_c\,l^2/r_H=2.36752$, and splits there exactly as the static pair does at $\hat q_c$. Beyond $a_\times$, or beyond $a_c$ when the pair collides, the late axisymmetric ringdown is a pure exponential. The time-domain evolution of Eq.~\eqref{eq:TD} confirms this: for $\hat q=0.3$ the signal at $a/l=1$ and $a/l=2$ in Fig.~\ref{fig:td}(b) decays without oscillation at rates $1.2306$ and $0.6757$, matching the HH values $1.230448$ and $0.675681$.

Measured in units of the temperature of the rotating string, the tower becomes regular. At $a/l=2.6$ the five lowest members of the neutral string sit at $\Gamma/2\pi T=1.157$, $2.216$, $3.287$, $4.365$ and $5.451$, so the spacing is close to one and decreases with $a$, as shown in Fig.~\ref{fig:rot}(c). For $\hat q=0.5$ at $a/l=2.5$ the same members sit at $1.275$, $2.406$, $3.511$, $4.591$ and $5.647$. A ladder with spacing close to the Matsubara value suggests that, at large boost, these modes are controlled by the horizon rather than by the AdS boundary.

\begin{figure*}[t]
\centering
\includegraphics[width=0.86\textwidth]{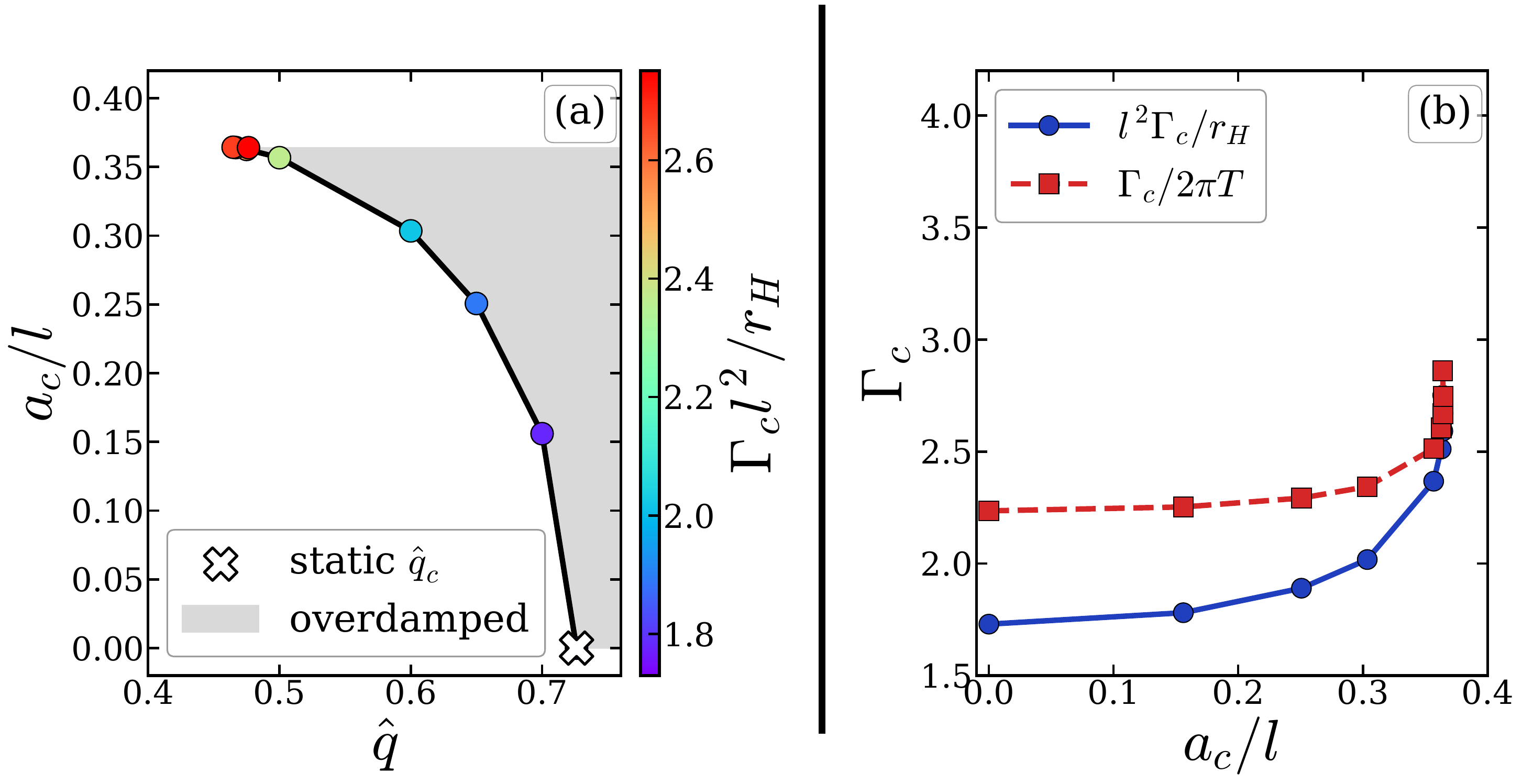}
\caption{Exceptional line of the axisymmetric ($m=0$, $k=0$) fundamental of the rotating RBS1 string. (a) Rotation $a_c$ at which the complex pair meets on the imaginary axis, as a function of the reduced charge, colored by the collision rate $\Gamma_c$. The line ends on the static exceptional point (cross) and folds near $\hat q\simeq0.465$. In the shaded region the pair is purely imaginary; the region above the fold was not mapped. (b) Collision rate along the line in units of $r_H/l^2$ and of $2\pi T=2\pi T_s/\gamma$.}
\label{fig:epline}
\end{figure*}

The two collisions are the ends of one curve. Solving the double-root conditions with $a$ and $\hat q$ as unknowns traces an exceptional line in the plane of charge and rotation, shown in Fig.~\ref{fig:epline}. It starts at $(\hat q_c,0)$, rises as $a_c/l\simeq0.96\,(\hat q_c-\hat q)^{1/2}$ near the static point, passes through $a_c/l=0.3035$ at $\hat q=0.6$ and $0.3567$ at $\hat q=0.5$, and folds at $\hat q\simeq0.465$, $a/l\simeq0.364$. Rotation therefore lowers the charge needed to overdamp the axisymmetric fundamental from $0.726$ to about $0.465$, although close to the fold the overdamped window in $a$ is narrow. Along the line the collision rate grows from $1.73$ to $2.67\,r_H/l^2$ at the fold, and in units of the temperature from $2.24$ to $2.75$. For the neutral string we followed the $m=0$ pair up to $a/l=3$ without finding a collision.

The complex branches split according to the sign of $m$. Counter-rotating modes ($m<0$) and the $m=0$ complex mode become more strongly damped as $a$ grows. The co-rotating $m=1$ mode instead becomes long-lived: its frequency approaches the superradiant threshold from above, with $\mathrm{Re}\,\omega-\Omega_H=0.071\,r_H/l^2$ for $\hat q=0$ and $0.057\,r_H/l^2$ for $\hat q=0.5$ at $a/l=3$, while its damping in units of the rotating temperature settles near $\mathrm{Im}\,\omega/2\pi T\simeq-1.14$ and $-1.26$, respectively. In the static frame this mode has $\mathrm{Re}\,\Omega=\gamma(\mathrm{Re}\,\omega-m\Omega_H)>0$, so it stays on the nonsuperradiant side, which is the statement that the boundary rotation of Eq.~\eqref{eq:OmegaH} never reaches the speed of light. The counter-rotating mode of the charged string behaves differently: at $\hat q=0.5$ its distance from the threshold falls to $0.04\,r_H/l^2$ near $a/l\simeq1.78$, where it is strongly damped, $\mathrm{Im}\,\omega\,l^2/r_H\simeq-3.95$, and where our continuation stops. We found $\mathrm{Im}\,\omega<0$ for every mode, charge and rotation we examined, for $0\le a/l\le3$, $|m|\le1$ and $\hat q\in\{0,0.5\}$, so the scalar sector of the rotating power-Maxwell string is linearly stable in this range.

\section{Consistency checks and scope}\label{isec7}

The results rest on checks of three kinds. The background was verified by direct substitution into the field equations, the wave equation by symbolic separation on the rotating metric, and the exact $\hat q=1$ spectrum by reinserting the terminating hypergeometric solutions into the radial equation, each in two independent symbolic calculations. Every number quoted in Tables~\ref{tab:bench} and \ref{tab:static} was obtained by at least two independent methods; the HH series and the collocation scheme agree to six or more digits, the two implementations of each method agree to ten or more digits, and the time domain agrees to three digits. The exceptional point was located both by counting imaginary roots and by solving the double-root conditions, and the asymptotic law of Sec.~\ref{isec5} reproduces the known neutral limit before being applied to the charged strings.

The scope of the statements is as follows. The field is a massless test scalar on a fixed background, so the spectrum describes the scalar sector only; electromagnetic perturbations couple to the NLED background and gravitational perturbations involve the full set of master equations, and neither is covered. The scalar decay rates bound nothing about the gravitational ringdown and carry no observational claim of their own. In holographic language the modes are the poles of the retarded correlator of a dimension-three scalar operator in a deformed thermal state. Our stability statement is linear and refers to the parameter ranges quantified above. The near-horizon estimate is tested only in the form that follows from the earlier treatment at $P=0$.

What is universal and what is model dependent can be stated compactly. The boost map, Eq.~\eqref{eq:boostmap}, holds for any metric function and any field content that separates in the static frame, and so does the statement that $\mathrm{Im}\,\omega/2\pi T$ depends on rotation only through $P^2$. The large-momentum exponent and phase depend only on the leading near-boundary power $\nu$, through Eq.~\eqref{eq:largek}, and the constants $\lambda_j$ are universal for a given $\nu$. The model enters through three numbers: the power $\nu=2-p=-4s/(1-2s)$, the coefficient $\hat q$, and the location $\hat q_c$ of the exceptional point, which has no analog in the neutral string.

\section{Conclusion}\label{isec8}

We revisited the scalar quasinormal spectrum of rotating AdS black strings sourced by a power-Maxwell field. The physical branch of the solution has negative exponent $s$, a charge term $-Qr^{p}$ with $p=2/(1-2s)$, and an algebraic horizon in the RBS1 case $s=-1/2$; the linear charge term is exact only for that value. Because rotation is a local boost, the rotating spectrum is fixed by the static one at a frequency-dependent momentum, and this exact map replaces the near-horizon estimate on which the stability of the model had been argued.

With three independent numerical methods, each implemented twice, benchmarked against the planar SAdS$_4$ spectrum and against an exact solution, we found that the NLED charge changes the scalar ringdown in two ways that a near-horizon treatment cannot capture. At $\hat q_c=0.72625663$ the least damped pair of modes meets on the imaginary axis, and the ringdown of strings with a larger charge-to-mass ratio is overdamped; the overtones follow at slightly larger charge, and at the endpoint where the mass term vanishes the spectrum is $\omega_n=-i(r_H/l^2)(n+1)(n+3)/[2(n+2)]$. At large transverse momentum the deviation from the light cone grows as $k^{1/(1-4s)}$ with a phase and a prefactor fixed by a one-dimensional eigenvalue problem, so the long-lived high-momentum modes of the neutral brane disappear. Rotation, for its part, continues the static exceptional point into a line in the plane of charge and rotation, which reaches down to $\hat q\simeq0.465$ before it folds, so that a rotating string overdamps its axisymmetric ringdown at a lower charge than a static one. It also adds a purely damped axisymmetric tower whose spacing approaches the Matsubara value in units of the rotating temperature, and it drives the co-rotating mode toward the superradiant threshold without crossing it, in line with the Hawking-Reall bound that the rotating string satisfies for every $a$.

The broader lesson is that in AdS black strings with matter fields that grow toward the boundary, the near-boundary behavior of the metric function is as important as the surface gravity, and the large-momentum law offers a direct diagnostic of it. The boost map also makes rotating strings a cheap laboratory, since every rotating result is a static computation at complex momentum.

In the near future, we plan to extend the analysis to electromagnetic perturbations, which couple to the power-Maxwell background and can probe the exceptional point in the vector channel, and to gravitational perturbations of the same strings. A second task is to compute the pseudospectrum near $\hat q_c$, where the coalescence of modes should make the spectrum most sensitive to small deformations, and to evaluate the retarded correlator of the dual operator across the overdamping transition.

\begin{acknowledgments}
\.{I}.S. is grateful to Eastern Mediterranean University (EMU), the Scientific and Technological Research Council of T\"urkiye (T\"UB\.{I}TAK), the Turkish consortium ANKOS and SCOAP3 for their support. \.{I}.S. also acknowledges the networking support of COST Actions CA18108, CA22113, CA21106, CA23130, CA23115 and CA21136.
\end{acknowledgments}

\section*{Data availability}
No observational data were used. The computational worksheets that produce the figures and tables can be supplied by the authors upon reasonable request.

\bibliographystyle{apsrev4-2}
\bibliography{ref}

\begin{thebibliography}{57}%
\makeatletter
\providecommand \@ifxundefined [1]{%
 \@ifx{#1\undefined}
}%
\providecommand \@ifnum [1]{%
 \ifnum #1\expandafter \@firstoftwo
 \else \expandafter \@secondoftwo
 \fi
}%
\providecommand \@ifx [1]{%
 \ifx #1\expandafter \@firstoftwo
 \else \expandafter \@secondoftwo
 \fi
}%
\providecommand \natexlab [1]{#1}%
\providecommand \enquote  [1]{``#1''}%
\providecommand \bibnamefont  [1]{#1}%
\providecommand \bibfnamefont [1]{#1}%
\providecommand \citenamefont [1]{#1}%
\providecommand \href@noop [0]{\@secondoftwo}%
\providecommand \href [0]{\begingroup \@sanitize@url \@href}%
\providecommand \@href[1]{\@@startlink{#1}\@@href}%
\providecommand \@@href[1]{\endgroup#1\@@endlink}%
\providecommand \@sanitize@url [0]{\catcode `\\12\catcode `\$12\catcode
  `\&12\catcode `\#12\catcode `\^12\catcode `\_12\catcode `\%12\relax}%
\providecommand \@@startlink[1]{}%
\providecommand \@@endlink[0]{}%
\providecommand \url  [0]{\begingroup\@sanitize@url \@url }%
\providecommand \@url [1]{\endgroup\@href {#1}{\urlprefix }}%
\providecommand \urlprefix  [0]{URL }%
\providecommand \Eprint [0]{\href }%
\providecommand \doibase [0]{https://doi.org/}%
\providecommand \selectlanguage [0]{\@gobble}%
\providecommand \bibinfo  [0]{\@secondoftwo}%
\providecommand \bibfield  [0]{\@secondoftwo}%
\providecommand \translation [1]{[#1]}%
\providecommand \BibitemOpen [0]{}%
\providecommand \bibitemStop [0]{}%
\providecommand \bibitemNoStop [0]{.\EOS\space}%
\providecommand \EOS [0]{\spacefactor3000\relax}%
\providecommand \BibitemShut  [1]{\csname bibitem#1\endcsname}%
\let\auto@bib@innerbib\@empty
\bibitem [{\citenamefont {Kokkotas}\ and\ \citenamefont
  {Schmidt}(1999)}]{Kokkotas:1999bd}%
  \BibitemOpen
  \bibfield  {author} {\bibinfo {author} {\bibfnamefont {K.~D.}\ \bibnamefont
  {Kokkotas}}\ and\ \bibinfo {author} {\bibfnamefont {B.~G.}\ \bibnamefont
  {Schmidt}},\ }\href {https://doi.org/10.12942/lrr-1999-2} {\bibfield
  {journal} {\bibinfo  {journal} {Living Rev. Relativ.}\ }\textbf {\bibinfo
  {volume} {2}},\ \bibinfo {pages} {2} (\bibinfo {year} {1999})},\ \Eprint
  {https://arxiv.org/abs/gr-qc/9909058} {gr-qc/9909058} \BibitemShut {NoStop}%
\bibitem [{\citenamefont {Berti}\ \emph {et~al.}(2009)\citenamefont {Berti},
  \citenamefont {Cardoso},\ and\ \citenamefont {Starinets}}]{Berti:2009kk}%
  \BibitemOpen
  \bibfield  {author} {\bibinfo {author} {\bibfnamefont {E.}~\bibnamefont
  {Berti}}, \bibinfo {author} {\bibfnamefont {V.}~\bibnamefont {Cardoso}},\
  and\ \bibinfo {author} {\bibfnamefont {A.~O.}\ \bibnamefont {Starinets}},\
  }\href {https://doi.org/10.1088/0264-9381/26/16/163001} {\bibfield  {journal}
  {\bibinfo  {journal} {Class. Quantum Grav.}\ }\textbf {\bibinfo {volume}
  {26}},\ \bibinfo {pages} {163001} (\bibinfo {year} {2009})}\BibitemShut
  {NoStop}%
\bibitem [{\citenamefont {Konoplya}\ and\ \citenamefont
  {Zhidenko}(2011)}]{Konoplya:2011qq}%
  \BibitemOpen
  \bibfield  {author} {\bibinfo {author} {\bibfnamefont {R.~A.}\ \bibnamefont
  {Konoplya}}\ and\ \bibinfo {author} {\bibfnamefont {A.}~\bibnamefont
  {Zhidenko}},\ }\href {https://doi.org/10.1103/RevModPhys.83.793} {\bibfield
  {journal} {\bibinfo  {journal} {Rev. Mod. Phys.}\ }\textbf {\bibinfo {volume}
  {83}},\ \bibinfo {pages} {793} (\bibinfo {year} {2011})},\ \Eprint
  {https://arxiv.org/abs/1102.4014} {1102.4014} \BibitemShut {NoStop}%
\bibitem [{\citenamefont {Birmingham}\ \emph {et~al.}(2002)\citenamefont
  {Birmingham}, \citenamefont {Sachs},\ and\ \citenamefont
  {Solodukhin}}]{Birmingham:2001pj}%
  \BibitemOpen
  \bibfield  {author} {\bibinfo {author} {\bibfnamefont {D.}~\bibnamefont
  {Birmingham}}, \bibinfo {author} {\bibfnamefont {I.}~\bibnamefont {Sachs}},\
  and\ \bibinfo {author} {\bibfnamefont {S.~N.}\ \bibnamefont {Solodukhin}},\
  }\href {https://doi.org/10.1103/PhysRevLett.88.151301} {\bibfield  {journal}
  {\bibinfo  {journal} {Phys. Rev. Lett.}\ }\textbf {\bibinfo {volume} {88}},\
  \bibinfo {pages} {151301} (\bibinfo {year} {2002})},\ \Eprint
  {https://arxiv.org/abs/hep-th/0112055} {hep-th/0112055} \BibitemShut
  {NoStop}%
\bibitem [{\citenamefont {Son}\ and\ \citenamefont
  {Starinets}(2002)}]{Son:2002sd}%
  \BibitemOpen
  \bibfield  {author} {\bibinfo {author} {\bibfnamefont {D.~T.}\ \bibnamefont
  {Son}}\ and\ \bibinfo {author} {\bibfnamefont {A.~O.}\ \bibnamefont
  {Starinets}},\ }\href {https://doi.org/10.1088/1126-6708/2002/09/042}
  {\bibfield  {journal} {\bibinfo  {journal} {J. High Energy Phys.}\ }\textbf
  {\bibinfo {volume} {09}},\ \bibinfo {pages} {042}},\ \Eprint
  {https://arxiv.org/abs/hep-th/0205051} {hep-th/0205051} \BibitemShut
  {NoStop}%
\bibitem [{\citenamefont {Kovtun}\ and\ \citenamefont
  {Starinets}(2005)}]{Kovtun:2005ev}%
  \BibitemOpen
  \bibfield  {author} {\bibinfo {author} {\bibfnamefont {P.~K.}\ \bibnamefont
  {Kovtun}}\ and\ \bibinfo {author} {\bibfnamefont {A.~O.}\ \bibnamefont
  {Starinets}},\ }\href {https://doi.org/10.1103/PhysRevD.72.086009} {\bibfield
   {journal} {\bibinfo  {journal} {Phys. Rev. D}\ }\textbf {\bibinfo {volume}
  {72}},\ \bibinfo {pages} {086009} (\bibinfo {year} {2005})},\ \Eprint
  {https://arxiv.org/abs/hep-th/0506184} {hep-th/0506184} \BibitemShut
  {NoStop}%
\bibitem [{\citenamefont {Horowitz}\ and\ \citenamefont
  {Hubeny}(2000)}]{Horowitz:1999jd}%
  \BibitemOpen
  \bibfield  {author} {\bibinfo {author} {\bibfnamefont {G.~T.}\ \bibnamefont
  {Horowitz}}\ and\ \bibinfo {author} {\bibfnamefont {V.~E.}\ \bibnamefont
  {Hubeny}},\ }\href {https://doi.org/10.1103/PhysRevD.62.024027} {\bibfield
  {journal} {\bibinfo  {journal} {Phys. Rev. D}\ }\textbf {\bibinfo {volume}
  {62}},\ \bibinfo {pages} {024027} (\bibinfo {year} {2000})},\ \Eprint
  {https://arxiv.org/abs/hep-th/9909056} {hep-th/9909056} \BibitemShut
  {NoStop}%
\bibitem [{\citenamefont {Cardoso}\ and\ \citenamefont
  {Lemos}(2001)}]{Cardoso:2001vs}%
  \BibitemOpen
  \bibfield  {author} {\bibinfo {author} {\bibfnamefont {V.}~\bibnamefont
  {Cardoso}}\ and\ \bibinfo {author} {\bibfnamefont {J.~P.~S.}\ \bibnamefont
  {Lemos}},\ }\href {https://doi.org/10.1088/0264-9381/18/23/319} {\bibfield
  {journal} {\bibinfo  {journal} {Class. Quantum Grav.}\ }\textbf {\bibinfo
  {volume} {18}},\ \bibinfo {pages} {5257} (\bibinfo {year} {2001})},\ \Eprint
  {https://arxiv.org/abs/gr-qc/0107098} {gr-qc/0107098} \BibitemShut {NoStop}%
\bibitem [{\citenamefont {Starinets}(2002)}]{Starinets:2002br}%
  \BibitemOpen
  \bibfield  {author} {\bibinfo {author} {\bibfnamefont {A.}~\bibnamefont
  {Starinets}},\ }\href {https://doi.org/10.1103/PhysRevD.66.124013} {\bibfield
   {journal} {\bibinfo  {journal} {Phys. Rev. D}\ }\textbf {\bibinfo {volume}
  {66}},\ \bibinfo {pages} {124013} (\bibinfo {year} {2002})}\BibitemShut
  {NoStop}%
\bibitem [{\citenamefont {Morgan}\ \emph
  {et~al.}(2009{\natexlab{a}})\citenamefont {Morgan}, \citenamefont {Cardoso},
  \citenamefont {Miranda}, \citenamefont {Molina},\ and\ \citenamefont
  {Zanchin}}]{Morgan:2009vg}%
  \BibitemOpen
  \bibfield  {author} {\bibinfo {author} {\bibfnamefont {J.}~\bibnamefont
  {Morgan}}, \bibinfo {author} {\bibfnamefont {V.}~\bibnamefont {Cardoso}},
  \bibinfo {author} {\bibfnamefont {A.~S.}\ \bibnamefont {Miranda}}, \bibinfo
  {author} {\bibfnamefont {C.}~\bibnamefont {Molina}},\ and\ \bibinfo {author}
  {\bibfnamefont {V.~T.}\ \bibnamefont {Zanchin}},\ }\href
  {https://doi.org/10.1088/1126-6708/2009/09/117} {\bibfield  {journal}
  {\bibinfo  {journal} {J. High Energy Phys.}\ }\textbf {\bibinfo {volume}
  {09}},\ \bibinfo {pages} {117}}\BibitemShut {NoStop}%
\bibitem [{\citenamefont {Morgan}\ \emph
  {et~al.}(2009{\natexlab{b}})\citenamefont {Morgan}, \citenamefont {Cardoso},
  \citenamefont {Miranda}, \citenamefont {Molina},\ and\ \citenamefont
  {Zanchin}}]{Morgan:2009pn}%
  \BibitemOpen
  \bibfield  {author} {\bibinfo {author} {\bibfnamefont {J.}~\bibnamefont
  {Morgan}}, \bibinfo {author} {\bibfnamefont {V.}~\bibnamefont {Cardoso}},
  \bibinfo {author} {\bibfnamefont {A.~S.}\ \bibnamefont {Miranda}}, \bibinfo
  {author} {\bibfnamefont {C.}~\bibnamefont {Molina}},\ and\ \bibinfo {author}
  {\bibfnamefont {V.~T.}\ \bibnamefont {Zanchin}},\ }\href
  {https://doi.org/10.1103/PhysRevD.80.024024} {\bibfield  {journal} {\bibinfo
  {journal} {Phys. Rev. D}\ }\textbf {\bibinfo {volume} {80}},\ \bibinfo
  {pages} {024024} (\bibinfo {year} {2009}{\natexlab{b}})}\BibitemShut
  {NoStop}%
\bibitem [{\citenamefont {Jansen}(2017)}]{Jansen:2017oag}%
  \BibitemOpen
  \bibfield  {author} {\bibinfo {author} {\bibfnamefont {A.}~\bibnamefont
  {Jansen}},\ }\href {https://doi.org/10.1140/epjp/i2017-11825-9} {\bibfield
  {journal} {\bibinfo  {journal} {Eur. Phys. J. Plus}\ }\textbf {\bibinfo
  {volume} {132}},\ \bibinfo {pages} {546} (\bibinfo {year}
  {2017})}\BibitemShut {NoStop}%
\bibitem [{\citenamefont {Grozdanov}\ and\ \citenamefont
  {Starinets}(2019)}]{Grozdanov:2019uhi}%
  \BibitemOpen
  \bibfield  {author} {\bibinfo {author} {\bibfnamefont {S.}~\bibnamefont
  {Grozdanov}}\ and\ \bibinfo {author} {\bibfnamefont {A.~O.}\ \bibnamefont
  {Starinets}},\ }\href {https://doi.org/10.1007/JHEP04(2019)080} {\bibfield
  {journal} {\bibinfo  {journal} {J. High Energy Phys.}\ }\textbf {\bibinfo
  {volume} {04}},\ \bibinfo {pages} {080}}\BibitemShut {NoStop}%
\bibitem [{\citenamefont {Hegde}\ \emph {et~al.}(2026)\citenamefont {Hegde},
  \citenamefont {Juric},\ and\ \citenamefont {Kumara}}]{Hegde:2026}%
  \BibitemOpen
  \bibfield  {author} {\bibinfo {author} {\bibfnamefont {K.}~\bibnamefont
  {Hegde}}, \bibinfo {author} {\bibfnamefont {T.}~\bibnamefont {Juric}},\ and\
  \bibinfo {author} {\bibfnamefont {A.~N.}\ \bibnamefont {Kumara}},\ }\href
  {https://doi.org/10.1007/JHEP07(2026)075} {\bibfield  {journal} {\bibinfo
  {journal} {J. High Energy Phys.}\ }\textbf {\bibinfo {volume} {07}},\
  \bibinfo {pages} {75}}\BibitemShut {NoStop}%
\bibitem [{\citenamefont {Jaramillo}\ \emph {et~al.}(2021)\citenamefont
  {Jaramillo}, \citenamefont {{Panosso Macedo}},\ and\ \citenamefont {{Al
  Sheikh}}}]{Jaramillo:2020tuu}%
  \BibitemOpen
  \bibfield  {author} {\bibinfo {author} {\bibfnamefont {J.~L.}\ \bibnamefont
  {Jaramillo}}, \bibinfo {author} {\bibfnamefont {R.}~\bibnamefont {{Panosso
  Macedo}}},\ and\ \bibinfo {author} {\bibfnamefont {L.}~\bibnamefont {{Al
  Sheikh}}},\ }\href {https://doi.org/10.1103/PhysRevX.11.031003} {\bibfield
  {journal} {\bibinfo  {journal} {Phys. Rev. X}\ }\textbf {\bibinfo {volume}
  {11}},\ \bibinfo {pages} {031003} (\bibinfo {year} {2021})},\ \Eprint
  {https://arxiv.org/abs/2004.06434} {2004.06434} \BibitemShut {NoStop}%
\bibitem [{\citenamefont {Arean}\ \emph {et~al.}(2023)\citenamefont {Arean},
  \citenamefont {Garcia-Farina},\ and\ \citenamefont
  {Landsteiner}}]{Arean:2023ejh}%
  \BibitemOpen
  \bibfield  {author} {\bibinfo {author} {\bibfnamefont {D.}~\bibnamefont
  {Arean}}, \bibinfo {author} {\bibfnamefont {D.}~\bibnamefont
  {Garcia-Farina}},\ and\ \bibinfo {author} {\bibfnamefont {K.}~\bibnamefont
  {Landsteiner}},\ }\href {https://doi.org/10.1007/JHEP12(2023)187} {\bibfield
  {journal} {\bibinfo  {journal} {J. High Energy Phys.}\ }\textbf {\bibinfo
  {volume} {12}},\ \bibinfo {pages} {187}}\BibitemShut {NoStop}%
\bibitem [{\citenamefont {Cownden}\ \emph {et~al.}(2024)\citenamefont
  {Cownden}, \citenamefont {Pantelidou},\ and\ \citenamefont
  {Zilhao}}]{Cownden:2023dam}%
  \BibitemOpen
  \bibfield  {author} {\bibinfo {author} {\bibfnamefont {B.}~\bibnamefont
  {Cownden}}, \bibinfo {author} {\bibfnamefont {C.}~\bibnamefont
  {Pantelidou}},\ and\ \bibinfo {author} {\bibfnamefont {M.}~\bibnamefont
  {Zilhao}},\ }\href {https://doi.org/10.1007/JHEP05(2024)202} {\bibfield
  {journal} {\bibinfo  {journal} {J. High Energy Phys.}\ }\textbf {\bibinfo
  {volume} {05}},\ \bibinfo {pages} {202}}\BibitemShut {NoStop}%
\bibitem [{\citenamefont {Garcia-Farina}\ \emph {et~al.}(2026)\citenamefont
  {Garcia-Farina}, \citenamefont {Landsteiner}, \citenamefont {Romeu},\ and\
  \citenamefont {Saura-Bastida}}]{GarciaFarina:2026}%
  \BibitemOpen
  \bibfield  {author} {\bibinfo {author} {\bibfnamefont {D.}~\bibnamefont
  {Garcia-Farina}}, \bibinfo {author} {\bibfnamefont {K.}~\bibnamefont
  {Landsteiner}}, \bibinfo {author} {\bibfnamefont {P.~G.}\ \bibnamefont
  {Romeu}},\ and\ \bibinfo {author} {\bibfnamefont {P.}~\bibnamefont
  {Saura-Bastida}},\ }\href {https://doi.org/10.1007/JHEP08(2026)186}
  {\bibfield  {journal} {\bibinfo  {journal} {J. High Energy Phys.}\ }\textbf
  {\bibinfo {volume} {08}},\ \bibinfo {pages} {186}}\BibitemShut {NoStop}%
\bibitem [{\citenamefont {Crump}\ and\ \citenamefont
  {Santos}(2026)}]{Crump:2026}%
  \BibitemOpen
  \bibfield  {author} {\bibinfo {author} {\bibfnamefont {J.~R.~V.}\
  \bibnamefont {Crump}}\ and\ \bibinfo {author} {\bibfnamefont {J.~E.}\
  \bibnamefont {Santos}},\ }\href {https://doi.org/10.1007/JHEP10(2026)044}
  {\bibfield  {journal} {\bibinfo  {journal} {J. High Energy Phys.}\ }\textbf
  {\bibinfo {volume} {10}},\ \bibinfo {pages} {44}}\BibitemShut {NoStop}%
\bibitem [{\citenamefont {Pan}\ \emph {et~al.}(2025)\citenamefont {Pan},
  \citenamefont {Yu},\ and\ \citenamefont {Ling}}]{Pan:2025}%
  \BibitemOpen
  \bibfield  {author} {\bibinfo {author} {\bibfnamefont {W.-B.}\ \bibnamefont
  {Pan}}, \bibinfo {author} {\bibfnamefont {Z.}~\bibnamefont {Yu}},\ and\
  \bibinfo {author} {\bibfnamefont {Y.}~\bibnamefont {Ling}},\ }\href
  {https://doi.org/10.1007/JHEP09(2025)147} {\bibfield  {journal} {\bibinfo
  {journal} {J. High Energy Phys.}\ }\textbf {\bibinfo {volume} {09}},\
  \bibinfo {pages} {147}}\BibitemShut {NoStop}%
\bibitem [{\citenamefont {He}\ \emph {et~al.}(2010)\citenamefont {He},
  \citenamefont {Wang}, \citenamefont {Cai},\ and\ \citenamefont
  {Lin}}]{He:2010zb}%
  \BibitemOpen
  \bibfield  {author} {\bibinfo {author} {\bibfnamefont {X.}~\bibnamefont
  {He}}, \bibinfo {author} {\bibfnamefont {B.}~\bibnamefont {Wang}}, \bibinfo
  {author} {\bibfnamefont {R.-G.}\ \bibnamefont {Cai}},\ and\ \bibinfo {author}
  {\bibfnamefont {C.-Y.}\ \bibnamefont {Lin}},\ }\href
  {https://doi.org/10.1016/j.physletb.2010.04.006} {\bibfield  {journal}
  {\bibinfo  {journal} {Phys. Lett. B}\ }\textbf {\bibinfo {volume} {688}},\
  \bibinfo {pages} {230} (\bibinfo {year} {2010})}\BibitemShut {NoStop}%
\bibitem [{\citenamefont {Liu}\ \emph {et~al.}(2014)\citenamefont {Liu},
  \citenamefont {Zou},\ and\ \citenamefont {Wang}}]{Liu:2014gvf}%
  \BibitemOpen
  \bibfield  {author} {\bibinfo {author} {\bibfnamefont {Y.}~\bibnamefont
  {Liu}}, \bibinfo {author} {\bibfnamefont {D.-C.}\ \bibnamefont {Zou}},\ and\
  \bibinfo {author} {\bibfnamefont {B.}~\bibnamefont {Wang}},\ }\href
  {https://doi.org/10.1007/JHEP09(2014)179} {\bibfield  {journal} {\bibinfo
  {journal} {J. High Energy Phys.}\ }\textbf {\bibinfo {volume} {09}},\
  \bibinfo {pages} {179}}\BibitemShut {NoStop}%
\bibitem [{\citenamefont {Guo}\ \emph {et~al.}(2024)\citenamefont {Guo},
  \citenamefont {Xie},\ and\ \citenamefont {Miao}}]{Guo:2024}%
  \BibitemOpen
  \bibfield  {author} {\bibinfo {author} {\bibfnamefont {Y.}~\bibnamefont
  {Guo}}, \bibinfo {author} {\bibfnamefont {H.}~\bibnamefont {Xie}},\ and\
  \bibinfo {author} {\bibfnamefont {Y.-G.}\ \bibnamefont {Miao}},\ }\href
  {https://doi.org/10.1016/j.physletb.2024.138801} {\bibfield  {journal}
  {\bibinfo  {journal} {Phys. Lett. B}\ }\textbf {\bibinfo {volume} {855}},\
  \bibinfo {pages} {138801} (\bibinfo {year} {2024})}\BibitemShut {NoStop}%
\bibitem [{\citenamefont {Lemos}(1995)}]{Lemos:1994xp}%
  \BibitemOpen
  \bibfield  {author} {\bibinfo {author} {\bibfnamefont {J.~P.~S.}\
  \bibnamefont {Lemos}},\ }\href {https://doi.org/10.1016/0370-2693(95)00533-Q}
  {\bibfield  {journal} {\bibinfo  {journal} {Phys. Lett. B}\ }\textbf
  {\bibinfo {volume} {353}},\ \bibinfo {pages} {46} (\bibinfo {year} {1995})},\
  \Eprint {https://arxiv.org/abs/gr-qc/9404041} {gr-qc/9404041} \BibitemShut
  {NoStop}%
\bibitem [{\citenamefont {Lemos}\ and\ \citenamefont
  {Zanchin}(1996)}]{Lemos:1995cm}%
  \BibitemOpen
  \bibfield  {author} {\bibinfo {author} {\bibfnamefont {J.~P.~S.}\
  \bibnamefont {Lemos}}\ and\ \bibinfo {author} {\bibfnamefont {V.~T.}\
  \bibnamefont {Zanchin}},\ }\href {https://doi.org/10.1103/PhysRevD.54.3840}
  {\bibfield  {journal} {\bibinfo  {journal} {Phys. Rev. D}\ }\textbf {\bibinfo
  {volume} {54}},\ \bibinfo {pages} {3840} (\bibinfo {year} {1996})},\ \Eprint
  {https://arxiv.org/abs/hep-th/9511188} {hep-th/9511188} \BibitemShut
  {NoStop}%
\bibitem [{\citenamefont {Morgan}\ \emph {et~al.}(2013)\citenamefont {Morgan},
  \citenamefont {Miranda},\ and\ \citenamefont {Zanchin}}]{Morgan:2013dv}%
  \BibitemOpen
  \bibfield  {author} {\bibinfo {author} {\bibfnamefont {J.}~\bibnamefont
  {Morgan}}, \bibinfo {author} {\bibfnamefont {A.~S.}\ \bibnamefont
  {Miranda}},\ and\ \bibinfo {author} {\bibfnamefont {V.~T.}\ \bibnamefont
  {Zanchin}},\ }\href {https://doi.org/10.1007/JHEP03(2013)169} {\bibfield
  {journal} {\bibinfo  {journal} {J. High Energy Phys.}\ }\textbf {\bibinfo
  {volume} {03}},\ \bibinfo {pages} {169}}\BibitemShut {NoStop}%
\bibitem [{\citenamefont {Mamani}\ \emph {et~al.}(2018)\citenamefont {Mamani},
  \citenamefont {Morgan}, \citenamefont {Miranda},\ and\ \citenamefont
  {Zanchin}}]{Mamani:2018qzl}%
  \BibitemOpen
  \bibfield  {author} {\bibinfo {author} {\bibfnamefont {L.~A.~H.}\
  \bibnamefont {Mamani}}, \bibinfo {author} {\bibfnamefont {J.}~\bibnamefont
  {Morgan}}, \bibinfo {author} {\bibfnamefont {A.~S.}\ \bibnamefont
  {Miranda}},\ and\ \bibinfo {author} {\bibfnamefont {V.~T.}\ \bibnamefont
  {Zanchin}},\ }\href {https://doi.org/10.1103/PhysRevD.98.026006} {\bibfield
  {journal} {\bibinfo  {journal} {Phys. Rev. D}\ }\textbf {\bibinfo {volume}
  {98}},\ \bibinfo {pages} {026006} (\bibinfo {year} {2018})}\BibitemShut
  {NoStop}%
\bibitem [{\citenamefont {Cardoso}\ and\ \citenamefont
  {Yoshida}(2005)}]{Cardoso:2005vk}%
  \BibitemOpen
  \bibfield  {author} {\bibinfo {author} {\bibfnamefont {V.}~\bibnamefont
  {Cardoso}}\ and\ \bibinfo {author} {\bibfnamefont {S.}~\bibnamefont
  {Yoshida}},\ }\href {https://doi.org/10.1088/1126-6708/2005/07/009}
  {\bibfield  {journal} {\bibinfo  {journal} {J. High Energy Phys.}\ }\textbf
  {\bibinfo {volume} {07}},\ \bibinfo {pages} {009}}\BibitemShut {NoStop}%
\bibitem [{\citenamefont {Gregory}\ and\ \citenamefont
  {Laflamme}(1993)}]{Gregory:1993vy}%
  \BibitemOpen
  \bibfield  {author} {\bibinfo {author} {\bibfnamefont {R.}~\bibnamefont
  {Gregory}}\ and\ \bibinfo {author} {\bibfnamefont {R.}~\bibnamefont
  {Laflamme}},\ }\href {https://doi.org/10.1103/PhysRevLett.70.2837} {\bibfield
   {journal} {\bibinfo  {journal} {Phys. Rev. Lett.}\ }\textbf {\bibinfo
  {volume} {70}},\ \bibinfo {pages} {2837} (\bibinfo {year} {1993})},\ \Eprint
  {https://arxiv.org/abs/hep-th/9301052} {hep-th/9301052} \BibitemShut
  {NoStop}%
\bibitem [{\citenamefont {Konoplya}\ \emph {et~al.}(2008)\citenamefont
  {Konoplya}, \citenamefont {Murata}, \citenamefont {Soda},\ and\ \citenamefont
  {Zhidenko}}]{Konoplya:2008yy}%
  \BibitemOpen
  \bibfield  {author} {\bibinfo {author} {\bibfnamefont {R.~A.}\ \bibnamefont
  {Konoplya}}, \bibinfo {author} {\bibfnamefont {K.}~\bibnamefont {Murata}},
  \bibinfo {author} {\bibfnamefont {J.}~\bibnamefont {Soda}},\ and\ \bibinfo
  {author} {\bibfnamefont {A.}~\bibnamefont {Zhidenko}},\ }\href
  {https://doi.org/10.1103/PhysRevD.78.084012} {\bibfield  {journal} {\bibinfo
  {journal} {Phys. Rev. D}\ }\textbf {\bibinfo {volume} {78}},\ \bibinfo
  {pages} {084012} (\bibinfo {year} {2008})}\BibitemShut {NoStop}%
\bibitem [{\citenamefont {Liu}\ and\ \citenamefont {Wang}(2008)}]{Liu:2008ds}%
  \BibitemOpen
  \bibfield  {author} {\bibinfo {author} {\bibfnamefont {L.}~\bibnamefont
  {Liu}}\ and\ \bibinfo {author} {\bibfnamefont {B.}~\bibnamefont {Wang}},\
  }\href {https://doi.org/10.1103/PhysRevD.78.064001} {\bibfield  {journal}
  {\bibinfo  {journal} {Phys. Rev. D}\ }\textbf {\bibinfo {volume} {78}},\
  \bibinfo {pages} {064001} (\bibinfo {year} {2008})}\BibitemShut {NoStop}%
\bibitem [{\citenamefont {Ponglertsakul}\ \emph {et~al.}(2018)\citenamefont
  {Ponglertsakul}, \citenamefont {Burikham},\ and\ \citenamefont
  {Tannukij}}]{Ponglertsakul:2018smo}%
  \BibitemOpen
  \bibfield  {author} {\bibinfo {author} {\bibfnamefont {S.}~\bibnamefont
  {Ponglertsakul}}, \bibinfo {author} {\bibfnamefont {P.}~\bibnamefont
  {Burikham}},\ and\ \bibinfo {author} {\bibfnamefont {L.}~\bibnamefont
  {Tannukij}},\ }\href {https://doi.org/10.1140/epjc/s10052-018-6057-9}
  {\bibfield  {journal} {\bibinfo  {journal} {Eur. Phys. J. C}\ }\textbf
  {\bibinfo {volume} {78}},\ \bibinfo {pages} {584} (\bibinfo {year}
  {2018})}\BibitemShut {NoStop}%
\bibitem [{\citenamefont {Lin}(2023)}]{Lin:2023}%
  \BibitemOpen
  \bibfield  {author} {\bibinfo {author} {\bibfnamefont {K.}~\bibnamefont
  {Lin}},\ }\href {https://doi.org/10.1103/PhysRevD.107.124002} {\bibfield
  {journal} {\bibinfo  {journal} {Phys. Rev. D}\ }\textbf {\bibinfo {volume}
  {107}},\ \bibinfo {pages} {124002} (\bibinfo {year} {2023})}\BibitemShut
  {NoStop}%
\bibitem [{\citenamefont {Ho}\ \emph {et~al.}(2022)\citenamefont {Ho},
  \citenamefont {Kim},\ and\ \citenamefont {Lee}}]{Ho:2022}%
  \BibitemOpen
  \bibfield  {author} {\bibinfo {author} {\bibfnamefont {J.}~\bibnamefont
  {Ho}}, \bibinfo {author} {\bibfnamefont {W.}~\bibnamefont {Kim}},\ and\
  \bibinfo {author} {\bibfnamefont {B.-H.}\ \bibnamefont {Lee}},\ }\href
  {https://doi.org/10.1007/JHEP08(2022)018} {\bibfield  {journal} {\bibinfo
  {journal} {J. High Energy Phys.}\ }\textbf {\bibinfo {volume} {08}},\
  \bibinfo {pages} {018}}\BibitemShut {NoStop}%
\bibitem [{\citenamefont {Emparan}\ \emph {et~al.}(2015)\citenamefont
  {Emparan}, \citenamefont {Suzuki},\ and\ \citenamefont
  {Tanabe}}]{Emparan:2014aba}%
  \BibitemOpen
  \bibfield  {author} {\bibinfo {author} {\bibfnamefont {R.}~\bibnamefont
  {Emparan}}, \bibinfo {author} {\bibfnamefont {R.}~\bibnamefont {Suzuki}},\
  and\ \bibinfo {author} {\bibfnamefont {K.}~\bibnamefont {Tanabe}},\ }\href
  {https://doi.org/10.1007/JHEP04(2015)085} {\bibfield  {journal} {\bibinfo
  {journal} {J. High Energy Phys.}\ }\textbf {\bibinfo {volume} {04}},\
  \bibinfo {pages} {085}}\BibitemShut {NoStop}%
\bibitem [{\citenamefont {Born}\ and\ \citenamefont
  {Infeld}(1934)}]{Born:1934gh}%
  \BibitemOpen
  \bibfield  {author} {\bibinfo {author} {\bibfnamefont {M.}~\bibnamefont
  {Born}}\ and\ \bibinfo {author} {\bibfnamefont {L.}~\bibnamefont {Infeld}},\
  }\href {https://doi.org/10.1098/rspa.1934.0059} {\bibfield  {journal}
  {\bibinfo  {journal} {Proc. R. Soc. Lond. A}\ }\textbf {\bibinfo {volume}
  {144}},\ \bibinfo {pages} {425} (\bibinfo {year} {1934})}\BibitemShut
  {NoStop}%
\bibitem [{\citenamefont {Hassaine}\ and\ \citenamefont
  {Martinez}(2007)}]{Hassaine:2007py}%
  \BibitemOpen
  \bibfield  {author} {\bibinfo {author} {\bibfnamefont {M.}~\bibnamefont
  {Hassaine}}\ and\ \bibinfo {author} {\bibfnamefont {C.}~\bibnamefont
  {Martinez}},\ }\href {https://doi.org/10.1103/PhysRevD.75.027502} {\bibfield
  {journal} {\bibinfo  {journal} {Phys. Rev. D}\ }\textbf {\bibinfo {volume}
  {75}},\ \bibinfo {pages} {027502} (\bibinfo {year} {2007})},\ \Eprint
  {https://arxiv.org/abs/hep-th/0701058} {hep-th/0701058} \BibitemShut
  {NoStop}%
\bibitem [{\citenamefont {Hendi}(2010)}]{Hendi:2010kv}%
  \BibitemOpen
  \bibfield  {author} {\bibinfo {author} {\bibfnamefont {S.~H.}\ \bibnamefont
  {Hendi}},\ }\href {https://doi.org/10.1103/PhysRevD.82.064040} {\bibfield
  {journal} {\bibinfo  {journal} {Phys. Rev. D}\ }\textbf {\bibinfo {volume}
  {82}},\ \bibinfo {pages} {064040} (\bibinfo {year} {2010})}\BibitemShut
  {NoStop}%
\bibitem [{\citenamefont {Hendi}\ and\ \citenamefont
  {Sheykhi}(2013)}]{Hendi:2013dwa}%
  \BibitemOpen
  \bibfield  {author} {\bibinfo {author} {\bibfnamefont {S.~H.}\ \bibnamefont
  {Hendi}}\ and\ \bibinfo {author} {\bibfnamefont {A.}~\bibnamefont
  {Sheykhi}},\ }\href {https://doi.org/10.1103/PhysRevD.88.044044} {\bibfield
  {journal} {\bibinfo  {journal} {Phys. Rev. D}\ }\textbf {\bibinfo {volume}
  {88}},\ \bibinfo {pages} {044044} (\bibinfo {year} {2013})}\BibitemShut
  {NoStop}%
\bibitem [{\citenamefont {Bakhtiarizadeh}\ and\ \citenamefont
  {Golchin}(2024)}]{Bakhtiarizadeh:2024}%
  \BibitemOpen
  \bibfield  {author} {\bibinfo {author} {\bibfnamefont {H.~R.}\ \bibnamefont
  {Bakhtiarizadeh}}\ and\ \bibinfo {author} {\bibfnamefont {H.}~\bibnamefont
  {Golchin}},\ }\href {https://doi.org/10.1088/1475-7516/2024/01/061}
  {\bibfield  {journal} {\bibinfo  {journal} {J. Cosmol. Astropart. Phys.}\
  }\textbf {\bibinfo {volume} {01}},\ \bibinfo {pages} {061}}\BibitemShut
  {NoStop}%
\bibitem [{\citenamefont {Barbosa}\ \emph {et~al.}(2025)\citenamefont
  {Barbosa}, \citenamefont {Ramos}, \citenamefont {Santos},\ and\ \citenamefont
  {Jr}}]{Barbosa:2025}%
  \BibitemOpen
  \bibfield  {author} {\bibinfo {author} {\bibfnamefont {L.~G.}\ \bibnamefont
  {Barbosa}}, \bibinfo {author} {\bibfnamefont {V.~H.~M.}\ \bibnamefont
  {Ramos}}, \bibinfo {author} {\bibfnamefont {L.~C.~N.}\ \bibnamefont
  {Santos}},\ and\ \bibinfo {author} {\bibfnamefont {C.~C.~B.}\ \bibnamefont
  {Jr}},\ }\href {https://doi.org/10.1088/1361-6382/add078} {\bibfield
  {journal} {\bibinfo  {journal} {Class. Quantum Grav.}\ }\textbf {\bibinfo
  {volume} {42}},\ \bibinfo {pages} {095016} (\bibinfo {year}
  {2025})}\BibitemShut {NoStop}%
\bibitem [{\citenamefont {Alencar}\ \emph {et~al.}(2026)\citenamefont
  {Alencar}, \citenamefont {Borralho}, \citenamefont {Crispim},\ and\
  \citenamefont {Cunha}}]{Alencar:2026reg}%
  \BibitemOpen
  \bibfield  {author} {\bibinfo {author} {\bibfnamefont {G.}~\bibnamefont
  {Alencar}}, \bibinfo {author} {\bibfnamefont {V.~H.~U.}\ \bibnamefont
  {Borralho}}, \bibinfo {author} {\bibfnamefont {T.~M.}\ \bibnamefont
  {Crispim}},\ and\ \bibinfo {author} {\bibfnamefont {M.~S.}\ \bibnamefont
  {Cunha}},\ }\href {https://doi.org/10.1088/1361-6382/ae8a0a} {\bibfield
  {journal} {\bibinfo  {journal} {Class. Quantum Grav.}\ }\textbf {\bibinfo
  {volume} {43}},\ \bibinfo {pages} {155011} (\bibinfo {year}
  {2026})}\BibitemShut {NoStop}%
\bibitem [{\citenamefont {Panotopoulos}\ and\ \citenamefont
  {Rincon}(2019)}]{Panotopoulos:2019}%
  \BibitemOpen
  \bibfield  {author} {\bibinfo {author} {\bibfnamefont {G.}~\bibnamefont
  {Panotopoulos}}\ and\ \bibinfo {author} {\bibfnamefont {A.}~\bibnamefont
  {Rincon}},\ }\href {https://doi.org/10.1142/S0218271819500160} {\bibfield
  {journal} {\bibinfo  {journal} {Int. J. Mod. Phys. D}\ }\textbf {\bibinfo
  {volume} {28}},\ \bibinfo {pages} {1950016} (\bibinfo {year}
  {2019})}\BibitemShut {NoStop}%
\bibitem [{\citenamefont {Liu}\ and\ \citenamefont {Wang}(2012)}]{Liu:2011bi}%
  \BibitemOpen
  \bibfield  {author} {\bibinfo {author} {\bibfnamefont {Y.}~\bibnamefont
  {Liu}}\ and\ \bibinfo {author} {\bibfnamefont {B.}~\bibnamefont {Wang}},\
  }\href {https://doi.org/10.1103/PhysRevD.85.046011} {\bibfield  {journal}
  {\bibinfo  {journal} {Phys. Rev. D}\ }\textbf {\bibinfo {volume} {85}},\
  \bibinfo {pages} {046011} (\bibinfo {year} {2012})}\BibitemShut {NoStop}%
\bibitem [{\citenamefont {Liu}\ and\ \citenamefont {Jing}(2012)}]{Liu:2012epm}%
  \BibitemOpen
  \bibfield  {author} {\bibinfo {author} {\bibfnamefont {Y.}~\bibnamefont
  {Liu}}\ and\ \bibinfo {author} {\bibfnamefont {J.-L.}\ \bibnamefont {Jing}},\
  }\href {https://doi.org/10.1088/0256-307X/29/1/010402} {\bibfield  {journal}
  {\bibinfo  {journal} {Chin. Phys. Lett.}\ }\textbf {\bibinfo {volume} {29}},\
  \bibinfo {pages} {010402} (\bibinfo {year} {2012})}\BibitemShut {NoStop}%
\bibitem [{\citenamefont {Gonzalez}\ and\ \citenamefont
  {Vasquez}(2016)}]{Gonzalez:2016}%
  \BibitemOpen
  \bibfield  {author} {\bibinfo {author} {\bibfnamefont {P.~A.}\ \bibnamefont
  {Gonzalez}}\ and\ \bibinfo {author} {\bibfnamefont {Y.}~\bibnamefont
  {Vasquez}},\ }\href {https://doi.org/10.1007/s10509-016-2819-8} {\bibfield
  {journal} {\bibinfo  {journal} {Astrophys. Space Sci.}\ }\textbf {\bibinfo
  {volume} {361}},\ \bibinfo {pages} {224} (\bibinfo {year}
  {2016})}\BibitemShut {NoStop}%
\bibitem [{\citenamefont {Aragon}\ \emph {et~al.}(2021)\citenamefont {Aragon},
  \citenamefont {Gonzalez}, \citenamefont {Saavedra},\ and\ \citenamefont
  {Vasquez}}]{Aragon:2021}%
  \BibitemOpen
  \bibfield  {author} {\bibinfo {author} {\bibfnamefont {A.}~\bibnamefont
  {Aragon}}, \bibinfo {author} {\bibfnamefont {P.~A.}\ \bibnamefont
  {Gonzalez}}, \bibinfo {author} {\bibfnamefont {J.}~\bibnamefont {Saavedra}},\
  and\ \bibinfo {author} {\bibfnamefont {Y.}~\bibnamefont {Vasquez}},\ }\href
  {https://doi.org/10.1007/s10714-021-02864-6} {\bibfield  {journal} {\bibinfo
  {journal} {Gen. Relativ. Gravit.}\ }\textbf {\bibinfo {volume} {53}},\
  \bibinfo {pages} {91} (\bibinfo {year} {2021})}\BibitemShut {NoStop}%
\bibitem [{\citenamefont {Mirekhtiary}\ \emph {et~al.}(2024)\citenamefont
  {Mirekhtiary}, \citenamefont {Abbasi}, \citenamefont {Hosseini},\ and\
  \citenamefont {Tulucu}}]{Mirekhtiary:2024rbs}%
  \BibitemOpen
  \bibfield  {author} {\bibinfo {author} {\bibfnamefont {F.~S.}\ \bibnamefont
  {Mirekhtiary}}, \bibinfo {author} {\bibfnamefont {A.}~\bibnamefont {Abbasi}},
  \bibinfo {author} {\bibfnamefont {K.}~\bibnamefont {Hosseini}},\ and\
  \bibinfo {author} {\bibfnamefont {F.}~\bibnamefont {Tulucu}},\ }\href
  {https://doi.org/10.1088/1402-4896/ad22cb} {\bibfield  {journal} {\bibinfo
  {journal} {Phys. Scr.}\ }\textbf {\bibinfo {volume} {99}},\ \bibinfo {pages}
  {035005} (\bibinfo {year} {2024})}\BibitemShut {NoStop}%
\bibitem [{\citenamefont {Mirekhtiary}\ and\ \citenamefont
  {Sakalli}(2023)}]{Mirekhtiary:2023qnm}%
  \BibitemOpen
  \bibfield  {author} {\bibinfo {author} {\bibfnamefont {F.~S.}\ \bibnamefont
  {Mirekhtiary}}\ and\ \bibinfo {author} {\bibfnamefont {I.}~\bibnamefont
  {Sakalli}},\ }\href {https://doi.org/10.1007/s12648-021-02262-3} {\bibfield
  {journal} {\bibinfo  {journal} {Indian J. Phys.}\ }\textbf {\bibinfo {volume}
  {97}},\ \bibinfo {pages} {1} (\bibinfo {year} {2023})}\BibitemShut {NoStop}%
\bibitem [{\citenamefont {Festuccia}\ and\ \citenamefont
  {Liu}(2008)}]{Festuccia:2008zx}%
  \BibitemOpen
  \bibfield  {author} {\bibinfo {author} {\bibfnamefont {G.}~\bibnamefont
  {Festuccia}}\ and\ \bibinfo {author} {\bibfnamefont {H.}~\bibnamefont
  {Liu}},\ }\href {https://arxiv.org/abs/0811.1033} {\bibfield  {journal}
  {\bibinfo  {journal} {arXiv:0811.1033}\ } (\bibinfo {year} {2008})},\ \Eprint
  {https://arxiv.org/abs/0811.1033} {0811.1033} \BibitemShut {NoStop}%
\bibitem [{\citenamefont {Hawking}\ and\ \citenamefont
  {Reall}(2000)}]{Hawking:1999dp}%
  \BibitemOpen
  \bibfield  {author} {\bibinfo {author} {\bibfnamefont {S.~W.}\ \bibnamefont
  {Hawking}}\ and\ \bibinfo {author} {\bibfnamefont {H.~S.}\ \bibnamefont
  {Reall}},\ }\href {https://doi.org/10.1103/PhysRevD.61.024014} {\bibfield
  {journal} {\bibinfo  {journal} {Phys. Rev. D}\ }\textbf {\bibinfo {volume}
  {61}},\ \bibinfo {pages} {024014} (\bibinfo {year} {2000})},\ \Eprint
  {https://arxiv.org/abs/hep-th/9908109} {hep-th/9908109} \BibitemShut
  {NoStop}%
\bibitem [{\citenamefont {Breitenlohner}\ and\ \citenamefont
  {Freedman}(1982)}]{Breitenlohner:1982jf}%
  \BibitemOpen
  \bibfield  {author} {\bibinfo {author} {\bibfnamefont {P.}~\bibnamefont
  {Breitenlohner}}\ and\ \bibinfo {author} {\bibfnamefont {D.~Z.}\ \bibnamefont
  {Freedman}},\ }\href {https://doi.org/10.1016/0003-4916(82)90116-6}
  {\bibfield  {journal} {\bibinfo  {journal} {Ann. Phys. (N.Y.)}\ }\textbf
  {\bibinfo {volume} {144}},\ \bibinfo {pages} {249} (\bibinfo {year}
  {1982})}\BibitemShut {NoStop}%
\bibitem [{\citenamefont {Zhidenko}(2006)}]{Zhidenko:2005mv}%
  \BibitemOpen
  \bibfield  {author} {\bibinfo {author} {\bibfnamefont {A.}~\bibnamefont
  {Zhidenko}},\ }\href {https://doi.org/10.1088/0264-9381/23/9/024} {\bibfield
  {journal} {\bibinfo  {journal} {Class. Quantum Grav.}\ }\textbf {\bibinfo
  {volume} {23}},\ \bibinfo {pages} {3155} (\bibinfo {year}
  {2006})}\BibitemShut {NoStop}%
\bibitem [{\citenamefont {Fernandes}\ \emph {et~al.}(2022)\citenamefont
  {Fernandes}, \citenamefont {Hilditch}, \citenamefont {Lemos},\ and\
  \citenamefont {Cardoso}}]{Fernandes:2021}%
  \BibitemOpen
  \bibfield  {author} {\bibinfo {author} {\bibfnamefont {V.~T.}\ \bibnamefont
  {Fernandes}}, \bibinfo {author} {\bibfnamefont {D.}~\bibnamefont {Hilditch}},
  \bibinfo {author} {\bibfnamefont {J.~P.~S.}\ \bibnamefont {Lemos}},\ and\
  \bibinfo {author} {\bibfnamefont {V.}~\bibnamefont {Cardoso}},\ }\href
  {https://doi.org/10.1103/PhysRevD.105.044017} {\bibfield  {journal} {\bibinfo
   {journal} {Phys. Rev. D}\ }\textbf {\bibinfo {volume} {105}},\ \bibinfo
  {pages} {044017} (\bibinfo {year} {2022})}\BibitemShut {NoStop}%
\bibitem [{\citenamefont {Gundlach}\ \emph {et~al.}(1994)\citenamefont
  {Gundlach}, \citenamefont {Price},\ and\ \citenamefont
  {Pullin}}]{Gundlach:1993tp}%
  \BibitemOpen
  \bibfield  {author} {\bibinfo {author} {\bibfnamefont {C.}~\bibnamefont
  {Gundlach}}, \bibinfo {author} {\bibfnamefont {R.~H.}\ \bibnamefont
  {Price}},\ and\ \bibinfo {author} {\bibfnamefont {J.}~\bibnamefont
  {Pullin}},\ }\href {https://doi.org/10.1103/PhysRevD.49.883} {\bibfield
  {journal} {\bibinfo  {journal} {Phys. Rev. D}\ }\textbf {\bibinfo {volume}
  {49}},\ \bibinfo {pages} {883} (\bibinfo {year} {1994})},\ \Eprint
  {https://arxiv.org/abs/gr-qc/9307009} {gr-qc/9307009} \BibitemShut {NoStop}%
\bibitem [{\citenamefont {Wang}\ \emph {et~al.}(2001)\citenamefont {Wang},
  \citenamefont {Molina},\ and\ \citenamefont {Abdalla}}]{Wang:2000gsa}%
  \BibitemOpen
  \bibfield  {author} {\bibinfo {author} {\bibfnamefont {B.}~\bibnamefont
  {Wang}}, \bibinfo {author} {\bibfnamefont {C.}~\bibnamefont {Molina}},\ and\
  \bibinfo {author} {\bibfnamefont {E.}~\bibnamefont {Abdalla}},\ }\href
  {https://doi.org/10.1103/PhysRevD.63.084001} {\bibfield  {journal} {\bibinfo
  {journal} {Phys. Rev. D}\ }\textbf {\bibinfo {volume} {63}},\ \bibinfo
  {pages} {084001} (\bibinfo {year} {2001})},\ \Eprint
  {https://arxiv.org/abs/hep-th/0005143} {hep-th/0005143} \BibitemShut
  {NoStop}%
\bibitem [{\citenamefont {Hua}\ and\ \citenamefont {Sarkar}(1990)}]{Hua:1990}%
  \BibitemOpen
  \bibfield  {author} {\bibinfo {author} {\bibfnamefont {Y.}~\bibnamefont
  {Hua}}\ and\ \bibinfo {author} {\bibfnamefont {T.~K.}\ \bibnamefont
  {Sarkar}},\ }\href {https://doi.org/10.1109/29.56027} {\bibfield  {journal}
  {\bibinfo  {journal} {IEEE Trans. Acoust. Speech Signal Process.}\ }\textbf
  {\bibinfo {volume} {38}},\ \bibinfo {pages} {814} (\bibinfo {year}
  {1990})}\BibitemShut {NoStop}%
\end{thebibliography}%

\end{document}